\documentclass[a4paper,fleqn,usenatbib, twocolumn]{aastex61}

\usepackage{newtxtext,newtxmath}
\usepackage[T1]{fontenc}
\usepackage{ae,aecompl}

\def\lesssim{\mathrel{\hbox{\rlap{\hbox{\lower4pt\hbox{$\sim$}}}\hbox{$<$}}}}
\def\gtrsim{\mathrel{\hbox{\rlap{\hbox{\lower4pt\hbox{$\sim$}}}\hbox{$>$}}}}

\def\mic{\,$\mu $m}

\def\apjs{ApJSup}

\def\pasp{PASP} 

\received{Dec 18, 2017}
\submitjournal{ApJS}

\shorttitle{The \textit{Herschel} -- ATLAS Data Release 2}
\shortauthors{Maddox et al.}

\begin{document}

\title{The \textit{Herschel}-ATLAS Data Release 2 Paper II: Catalogues of
       far-infrared and sub-millimetre sources in the fields at the south
       and north Galactic Poles}

\correspondingauthor{S. J. Maddox}
\email{maddoxs@cardiff.ac.uk}

\author[0000-0001-5549-195X]{S.J. Maddox} \affiliation{School of
  Physics and Astronomy, Cardiff University, The Parade, Cardiff CF24
  3AA, UK.}  \affiliation{Institute for Astronomy, The University of
  Edinburgh, Royal Observatory, Blackford Hill, Edinburgh, EH9 3HJ,
  UK.}

\author{E. Valiante} \affiliation{School of Physics and Astronomy,
  Cardiff University, The Parade, Cardiff CF24 3AA, UK.}

\author{P. Cigan} \affiliation{School of Physics and Astronomy,
  Cardiff University, The Parade, Cardiff CF24 3AA, UK.}

\author{L. Dunne} \affiliation{School of Physics and Astronomy,
  Cardiff University, The Parade, Cardiff CF24 3AA, UK.}
\affiliation{Institute for Astronomy, The University of Edinburgh,
  Royal Observatory, Blackford Hill, Edinburgh, EH9 3HJ, UK.}

\author{S. Eales} \affiliation{School of Physics and Astronomy,
  Cardiff University, The Parade, Cardiff CF24 3AA, UK.}

\author{M. W. L. Smith} \affiliation{School of Physics and Astronomy,
  Cardiff University, The Parade, Cardiff CF24 3AA, UK.}

\author{S. Dye} \affiliation{School of Physics and Astronomy,
  University of Nottingham, University Park, Nottingham, NG7 2RD, UK.}

\author{C. Furlanetto} \affiliation{School of Physics and Astronomy,
  University of Nottingham, University Park, Nottingham, NG7 2RD, UK.}
\affiliation{Departamento de F\'isica, Universidade Federale do Rio
  Grande do Sul., Av. Bento Gon\c{c}alves, 9500, 91501-970, Porto Algres,
  RS Brazil}

\author{E. Ibar} \affiliation{Instituto de F\'isica y Astronom\'ia,
  Universidad de Valpara\'iso, Avda. Gran Breta\~na 1111,
  Valpara\'iso, Chile.}

\author{ G. de Zotti} \affiliation{INAF-Osservatorio Astronomico di
  Padova, Vicolo dell'Osservatorio 5, I-35122 Padova, Italy}

\author{J. S. Millard} \affiliation{School of Physics and Astronomy,
  Cardiff University, The Parade, Cardiff CF24 3AA, UK.}

\author{N. Bourne} \affiliation{Institute for Astronomy, The
  University of Edinburgh, Royal Observatory, Blackford Hill,
  Edinburgh, EH9 3HJ, UK.}

\author{H. L. Gomez} \affiliation{School of Physics and Astronomy,
  University of Nottingham, University Park, Nottingham, NG7 2RD, UK.}

\author{R. J. Ivison} \affiliation{European Southern Observatory,
  Karl-Schwarzschild-Strasse 2, 85748, Garching, Germany}
\affiliation{Institute for Astronomy, The University of Edinburgh,
  Royal Observatory, Blackford Hill, Edinburgh, EH9 3HJ, UK.}

\author{ D. Scott} \affiliation{Department of Physics \& Astronomy,
  University of British Columbia, 6224 Agricultural Road, Vancouver,
  BC V6T 1Z1, Canada}

\author{I. Valtchanov} \affiliation{Telespazio Vega UK for ESA,
  European Space Astronomy Centre, Operations Department, 28691
  Villanueva de la Ca\~nada, Spain}


\begin{abstract}

  The {\it Herschel} Astrophysical Terahertz Large Area Survey
  (H-ATLAS) is a survey of 660 deg$^2$ with the PACS and SPIRE cameras
  in five photometric bands: 100, 160, 250, 350 and 500\mic.  This is
  the second of three papers describing the data release for the large
  fields at the south and north Galactic poles (NGP and SGP).  In this
  paper we describe the catalogues of far-infrared and submillimetre
  sources for the NGP and SGP, which cover 177.1 deg$^2$ and 303.4
  deg$^2$, respectively.  The catalogues contain 118,908 sources for
  the NGP field and 193,527 sources for the SGP field detected at more
  than 4$\sigma$ significance in any of the 250, 350 or 500\mic\
  bands. The source detection is based on the 250\mic\ map, and we
  present photometry in all five bands for each source, including
  aperture photometry for sources known to be extended.  The rms
  positional accuracy for the faintest sources is about 2.4 arc
  seconds in both right ascension and declination.  We present a
  statistical analysis of the catalogues and discuss the practical
  issues -- completeness, reliability, flux boosting, accuracy of
  positions, accuracy of flux measurements -- necessary to use the
  catalogues for astronomical projects.
\end{abstract}


\section{Introduction}

This is the second of three papers describing the second major data
release of the {\it Herschel} Astrophysical Terahertz Large Area
Survey (the {\it Herschel} ATLAS or H-ATLAS), the largest single key
project carried out in open time with the {\it Herschel Space
  Observatory}\footnote{{\it Herschel} is an ESA space observatory with
  science instruments provided by European-led Principal Investigator
  consortia and with important participation from NASA} (Pilbratt et
al.  2010).  The H-ATLAS is a survey of approximately 660 deg$^2$ of
sky in five photometric bands: 100, 160, 250, 350 and 500\mic\
(Eales et al.  2010).  Although the original goal of the survey was to
study dust, and the newly formed stars hidden by dust, in galaxies in
the nearby ($z<0.4$) universe (Dunne et al. 2011, Eales et al. 2018),
in practice the exceptional sensitivity of {\it Herschel}, aided by
the large negative {\it k}-correction at submillimetre wavelengths
(Franceschini et al. 1991), has meant that the median redshift of the
sources detected in the survey is approximately 1 (Pearson et
al. 2013), and our source catalogues include sources up to a redshift
of at least 6 (Fudamoto et al. 2017; Zavala et al. 2017).

The five H-ATLAS fields were selected to be areas with relatively
little emission from dust in the Milky Way, as judged from the IRAS
100\mic\ images (Neugebauer 1984), and with a large amount of data
in other wavebands. In 2010 for the Science Demonstration Phase (SDP)
of {\it Herschel}, we provided the data products for one 16 deg$^2$
field in the GAMA 9-hour field (Ibar et al. 2010; Pascale et al. 2011;
Rigby et al. 2011; Smith et al. 2011).  In our first large data
release (DR1), we released the data products for three fields on the
celestial equator centred at R.A. approximately 9, 12 and 15 hours
(Valiante et al. 2016, hereafter V16; Bourne et al.  2016), covering a total area of
161 deg$^2$.  These data products included the {\it Herschel} images
in all five bands, a catalogue of the 120,230 sources detected in
these images and of the 44,835 optical counterparts to these sources.

\begin{figure} 
\includegraphics[scale=0.9, trim=0.0mm 5.0mm 0.0mm 0.0mm, clip=True]{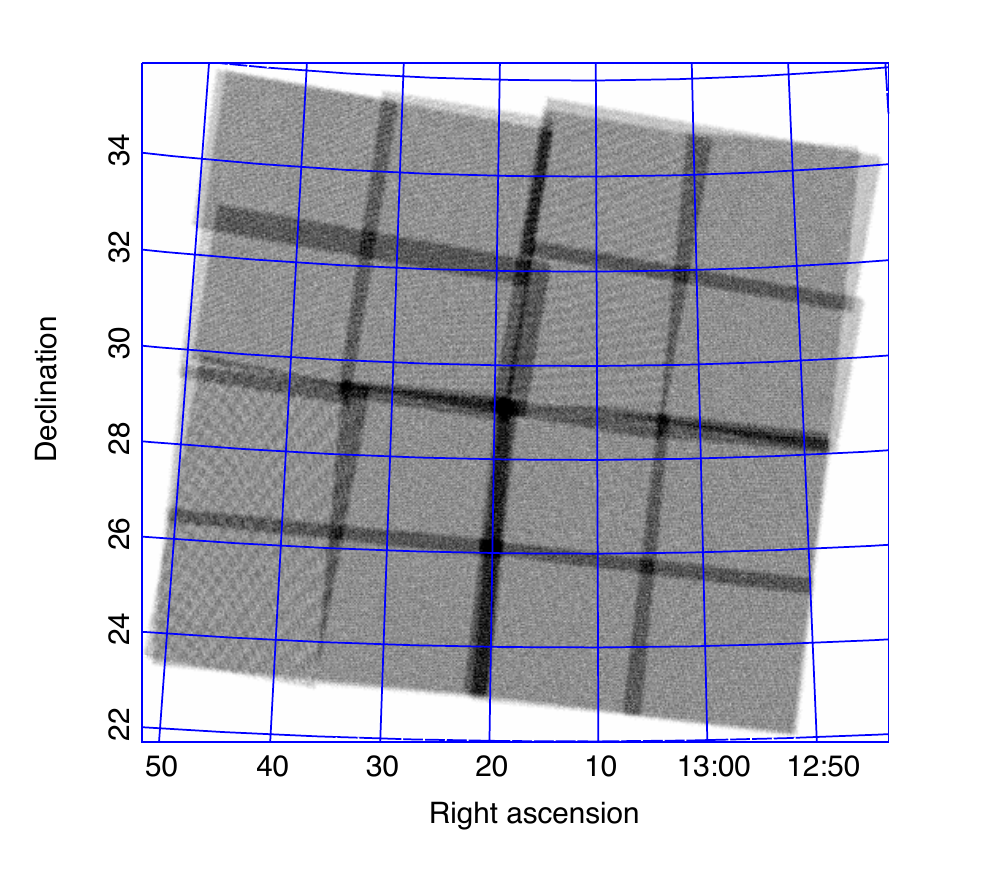}
\caption{ \protect\label{fig_ngp} The coverage map for 250\mic\
  observations of the NGP field.  The map shows the number of samples
  from the bolometer timelines contributing to each map pixel, which
  ranges from 1 to 43, with the median value being 10.  The range of
  the grayscale is from 0 samples (white) to 27 samples (black). }
\end{figure}

\begin{figure*} 
\includegraphics[scale=1., trim=0.0mm 5.0mm 0.0mm 5.0mm, clip=True]{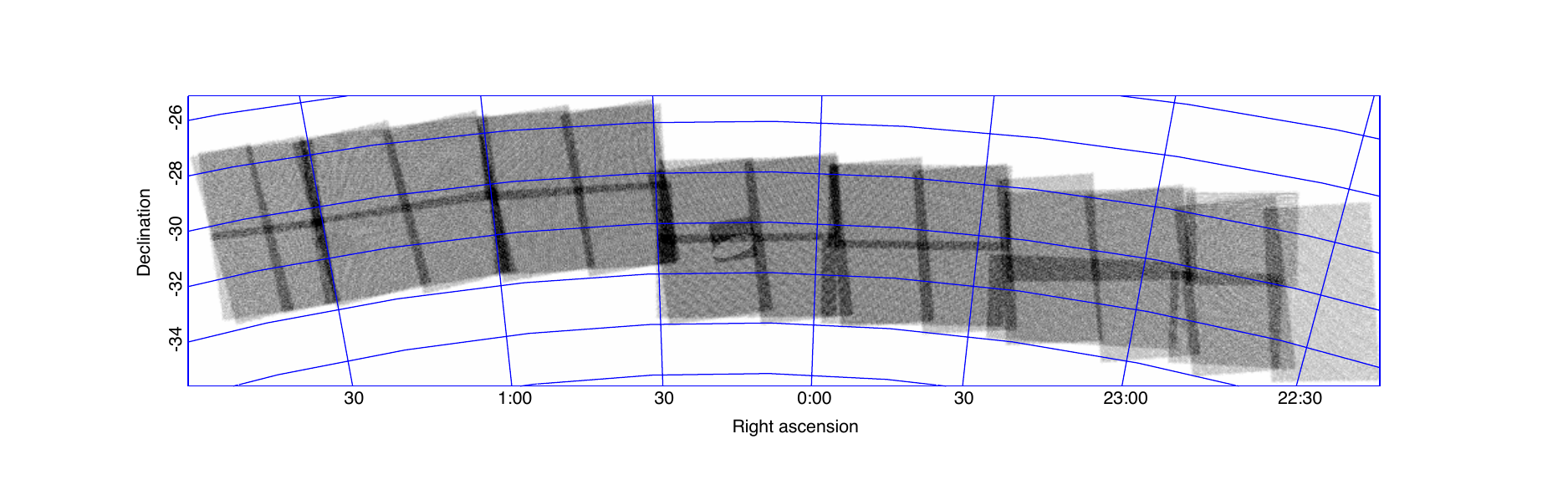}
\caption{ \protect\label{fig_sgp} The coverage map for 250\mic\
  observations of the SGP field.  The map shows the number of samples
  from the bolometer timelines contributing to each map pixel, which
  ranges from 1 to 36, with the median value being 9.  The range of
  the grayscale is from 0 samples (white) to 21 samples (black). }
\end{figure*}

Our second data release is for the two larger fields at the north and
south Galactic poles (NGP and SGP). The NGP field is centred
approximately at a right ascension of
$13^{\mathrm{h}}\ 18^{\mathrm{m}}$ and a declination of
$+29^{\circ}\ 13^{\prime}$ (J2000) and has an area of 180.1~deg$^2$.
The NGP field is a roughly square region (Fig.~\ref{fig_ngp}) and,
among many other interesting known extra-galactic objects, includes the
Coma Cluster.  The SGP field is centred approximately at a right
ascension of $ 0^{\mathrm{h}}\ 6^{\mathrm{m}}$ and a declination of
$-32^{\circ} \ 44^{\prime}$ (J2000) and has an area of
317.6~deg$^2$. The SGP field is elongated in right ascension
(Fig.~\ref{fig_sgp}).  Smith et al. (2017, hereafter S17) provide a
comprehensive list of the multi-wavelength data that exist for these
fields.

Our data release for these fields is described in three papers. In the
first paper (S17), we present the images of these fields, including a
description of how these images can be used by the astronomical
community for a variety of scientific projects.  In this paper, we
describe the production and properties of the catalogues of
far-infrared and submillimetre sources detected in these images. A
third paper (Furlanetto et al. 2017, hereafter F17) describes a search
for the optical/near-infrared counterparts to the {\it Herschel}
sources in the NGP field and the resulting multi-wavelength catalogue.
The catalogues described in this paper can be obtained from the
H-ATLAS website (\url{http://www.h-atlas.org}).

The arrangement of this paper is as follows. Section 2 describes the
maps and masks used to define the catalogues. Section 3 describes the
detection of the sources. Section 4 describes the photometry of the
sources. Section 5 describes the catalogues and their properties.
Finally Section 6 gives a summary of the paper.

\section{Maps, Coverage and Masks}

A detailed description of the processing necessary to produce maps
from the {\it Herschel} raw data is presented in S17. The resulting
maps have pixel sizes 3, 4, 6, 8 and 12 arcsec for 100, 160, 250,
350 and 500 $\mu$m, respectively. We chose these to
  optimally sample the PSF in each band, given that the FWHM of the PSF is
  11.4, 13.7, 17.8, 24 and 35.2 arcsec at 100, 160, 250, 350 and
  500\mic\ respectively.  Note that these are different to the
canonical pixel sizes used for maps in the {\it Herschel} Science
Archive, which use 3.2, 3.2 6, 10 and 14 arcsec respectively.
The maps made with the PACS camera (100 and 160\mic, Poglitsch et
al. 2010) have units of Jy per pixel. The maps made with the SPIRE
camera (250, 350 and 500\mic, Griffin et al. 2010) have units of Jy
per beam.  The beam areas at 250, 350 and 500\mic\ are 469, 831 and
1804 square arcsec, respectively (Valtchanov 2017).  The noise on
the images is a combination of instrumental noise and the confusion
noise from sources that are too faint to be detected individually.
S17 describes a detailed analysis of the noise properties of the
images.

\begin{deluxetable}{lrrrrrrrr}[h]

  \tablecaption{Area of the survey data  in deg$^2$, as
    a function of the number of {\it Herschel} observations
    ($N_{\mathrm{scan}}$). The entries with `total' show all of the
    observed area. The entries with `+mask' are the areas within the
    mask used to define the catalogue. \label{tab_areas} }
  \tablecolumns{9} \tablewidth{0pt} \tabletypesize{\scriptsize}
  \tablehead{ &  \multicolumn{7}{c}{$N_{\mathrm{scan}}$} \\
    & \colhead{1} & \colhead{2} & \colhead{3} & \colhead{4} &
    \colhead{5} & \colhead{6} & \colhead{7} & \colhead{total}}
  \startdata
NGP total   &  8.3 &139.3 & 26.2 &   5.7 &0.4 & 0.2 &0.1 &180.1 \\
NGP+mask    &  5.5 &139.1 & 26.2 &   5.7 &0.4 & 0.2 &0.1 &177.1\\
SGP total   & 43.0 &210.2 & 52.2 &  11.1 &0.7 & 0.2 &  0 &317.6\\
SGP+mask    & 30.5 &208.7 & 52.1 &  11.1 &0.7 & 0.2 &  0 &303.4\\
\enddata
\end{deluxetable}

\begin{figure} 

\includegraphics[scale=0.4, trim=0.0mm 35.0mm 0.0mm 25.0mm, clip=True]{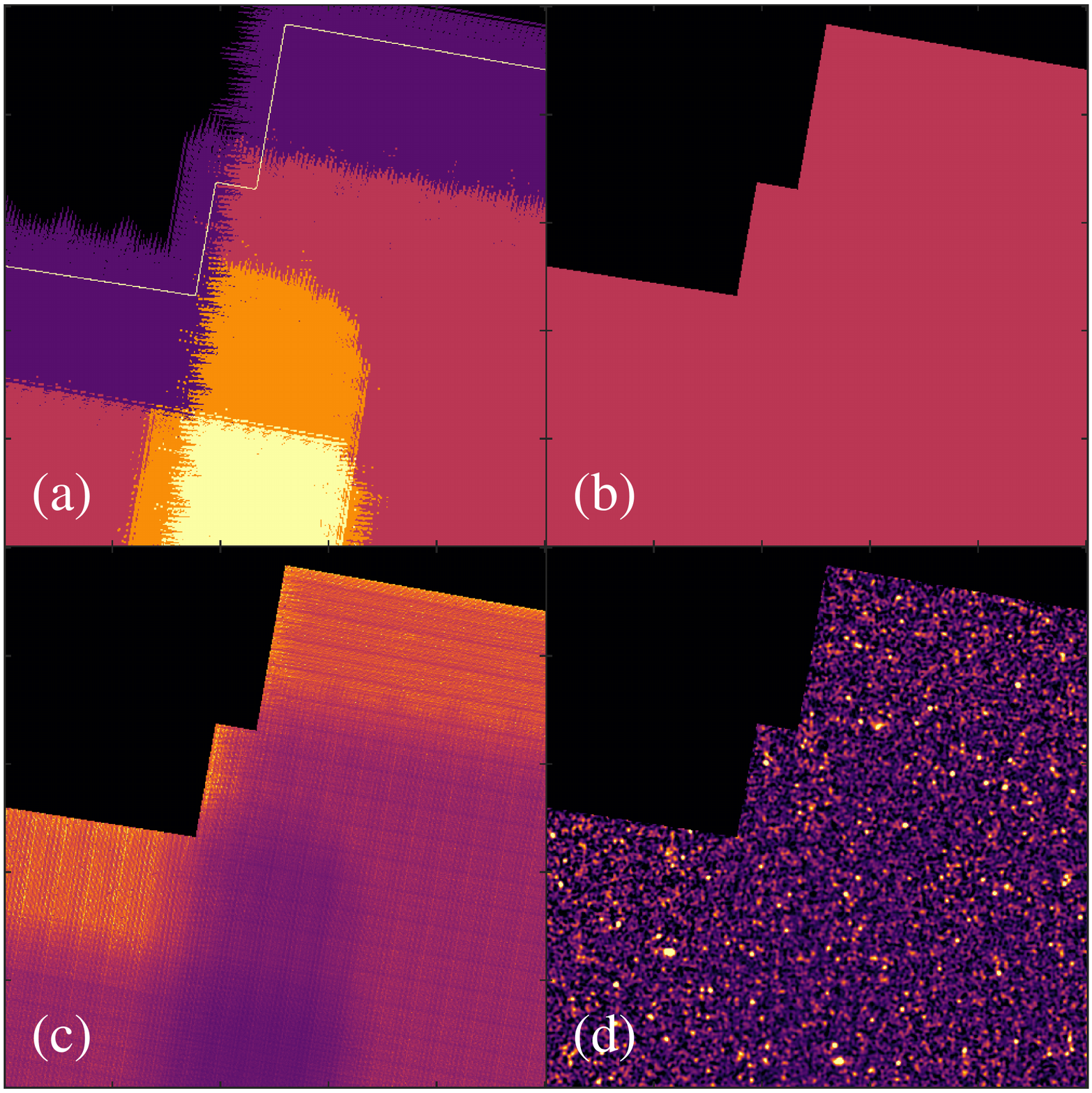}
\caption{\protect\label{fig_mask} (a) Coverage map for a
  $50'\times 50'$ region of the 250\mic\ observations, chosen from the
  NGP field to show the complicated variation in some areas of the
  survey.  The map shows the number of scans covering each pixel,
  ranging from 1 to 4. The line shows the edge of the masked region
  retained for the final data. (b) The mask corresponding to the same
  region. (c) The instrumental noise per pixel for the same
  region. The average noise varies between regions depending on the
  number of observing scans: roughly 14mJy, 10mJy, 8mJy and 7mJy for
  the 1, 2, 3 and 4 $N_{\mathrm{scan}}$ regions respectively. Also
  note that the noise varies significantly within regions where the
  number of scans in constant. This is due to the variation in the
  number of detector passes across a single scan of the instrument,
  and the pixel position relative to the time samples along the scan
  direction. (d) The 250\mic\ data for the same region.  }
\end{figure}

  The boundary of the mapping data is set by the coverage of the
  scan lines of the instrument, and so is very ragged, as shown in
  Fig.~\ref{fig_mask}(a). We define a simple mask to set a clear
  boundary for the data used in the catalogues; this is mostly
  restricted to the areas with more than one {\it Herschel}
  observation, but does include some areas with only one scan, as can
  be see in Fig.~\ref{fig_mask}(b).  The mask reduces the area covered
  by the catalogues to 177.1~deg$^2$ and 303.4~deg$^2$ for the NGP and
  SGP, respectively. The area covered by the NGP and SGP fields is
  listed as a function of the number of observations ($N_{\mathrm{scan}}$) in
  Table~\ref{tab_areas}. Within regions where the number of scans is
  constant, the mean noise is constant, but the noise varies
  significantly from pixel to pixel, as can be seen in
  Fig.~\ref{fig_mask}(c).  This is because the number of detector
  passes that contribute to a pixel depends on the pixel position
  relative to the detectors across the scan direction, and also the
  position relative to the time samples along the scan direction.
  
  The SPIRE and PACS photometers are offset by 21 arcmin, which
  creates regions around the borders of the survey that are covered by
  only one of the two photometers. As in previous data releases we
  restrict our catalogues to the area covered by the 250\mic\ maps, so
  there are some sources that do not have coverage in the PACS 100 and
  160\mic\ bands. 

\section{Source detection} 

\subsection{Background subtraction} 

Before attempting to detect sources in the maps, we first subtracted a
smoothly varying ``sky'' level to remove the foreground emission from
dust in our galaxy, so-called ``cirrus emission'', and also the
emission from clustered extra-galactic sources fainter than our
detection limit. We used the {\tt nebuliser} function, a programme
produced by the Cambridge Astronomy Survey Unit to estimate and
subtract the sky level on astronomical
images\footnote{\url{http://casu.ast.cam.ac.uk/surveys-projects/software-release/background-filtering}}.
The algorithm first applies a 2-D moving box-car median filter to
estimate the local sky level for each pixel, and then applies a 2-D
moving box-car mean filter to slightly smooth the resulting sky map.

The choice of the filter scale used in {\tt nebuliser} is quite
critical, since it must be small enough for {\tt nebuliser} to remove
small-scale patches of cirrus emission but not so small that the flux
from large galaxies is reduced.  In practice, for the SPIRE maps we
found that a median filter scale of 30 pixels (3 arcmin in the
250\mic\ band) followed by a linear filter scale of 15 pixels was an
acceptable combination.

We tested whether this filtering scale reduced the flux density of
extended extra-galactic sources by creating simulated maps, placing
artificial extended sources on these maps, and then measuring the flux
densities of these sources after the application of {\tt nebuliser}.
Since the nearby extended galaxies detected by {\it Herschel} are
mostly spiral galaxies, we used truncated exponential profiles for the
artificial sources, and convolved these with the SPIRE point-spread
function.  Previous surveys have found that the observed extent of FIR
emission is quite similar to the optical (Hunt et al. 2015, Smith et
al. 2012), and the widely used D25 optical diameters for galaxies are
roughly equivalent to a distance of five scale lengths from the centre
of a galaxy, so we truncated the profiles of our artificial
submillimetre sources at five scale lengths. At this radius the
profile contains 96\% of a non-truncated exponential; extending to six
scale lengths would increase this to 98\%, only a 2\% change, so the
exact truncation radius is not critical. The resulting diameters
ranged from 24 to 192 arcsec.  Since the diameters are much larger
than the PSF in all of the SPIRE bands, the results will be similar
for all bands.  The simulations showed that significant flux is lost
only for sources that have diameters larger than $~3$ arcmin, and even
for sources above this size, the flux loss is $\lesssim 10\%$.

Note there are only 12 galaxies with diameter larger than 3 arc
minutes in the survey: three in the NGP and nine in the SGP. We have
made no attempt to correct for any filtering-related flux losses for
these galaxies, and recommend users make their own measurements based
on the non-filtered maps if more precise extended photometry is
required.

We note that the application of {\tt nebuliser} will change the
clustering statistics of extra-galactic sources.  Apart from the
foreground cirrus emission, {\tt nebuliser} removes the background
produced by the sources that are too faint to be detected
individually. This background varies because of the clustering of
these faint sources.  A source catalogue made without any background
subtraction will include more sources where this background is high as
a result of clusters of these faint sources, and so the clustering of
the sources in such a catalogue will be stronger than in a catalogue
produced from an image in which this background emission has been
removed. An investigation of the clustering in the H-ATLAS catalogues,
which includes an analysis of the effect of the subtraction of this
background, will be presented by Amvrosiadis et al. (in preparation).

For the PACS maps, the $1/f$-noise from the instrument is much larger than
for SPIRE, making the foreground cirrus emission and the background
emission from faint galaxies difficult to detect. Since we could not
clearly detect the foreground/background emission on smaller
scales, we used a {\tt nebuliser} scale of 5 arcminutes.

The raw maps from the SPIRE pipeline have a mean of zero, but the
output maps from {\tt nebuliser} have a modal pixel value that is
zero. For the SPIRE bands, the instrumental noise is low enough that
the flux distribution of detected sources skews the pixel distribution
to positive values so the mean is slightly positive (1.0, 1.0 and 0.6
mJy/beam at 250, 350 and 500\mic).  The PACS detector is less
sensitive and less stable than SPIRE, and so the instrumental noise
dominates over the confusion noise and the pixel distribution is close
to Gaussian; the mean of the {\tt nebulised} PACS maps are very close
to zero (0.016\,MJy\,sr$^{-1}$, for both the 100 and 160\mic\ maps).

\subsection{Source Detection}

In this section we describe the method used to find the
sources on the images. Additional details are given in V16.
Sources were detected using the {\tt MADX} algorithm (Maddox et al in prep)
applied to the SPIRE maps.  {\tt MADX} creates maps of the signal-to-noise
ratio and identifies sources by finding peaks in the signal to noise. The
detection and measurement of fluxes is optimised by using a matched
filter that is applied to both the signal map and the noise map. 

The SPIRE instrumental noise maps are created from the number of
detector passes and the estimated instrumental noise per pass,
$\sigma_{\mathrm{inst}} /\sqrt{N_ {\mathrm{sample}}}$, as described in S17 and V16.

 
Since the noise consists of both instrumental noise and
confusion noise from the background of undetected sources, we follow
the approach of Chapin et al (2011) to calculate the optimal matched
filter in each of the three SPIRE bands. Details of the estimation and
form of the matched filter are discussed in V16.  The resulting
matched filters are slightly more compact than the corresponding PSFs,
and have slightly negative regions outside the FWHM.  

In the first step of the source detection, peak pixels which have
values $>2.5\sigma$ in the filtered 250-\micron\ map are considered as
potential sources.  We use the 250-\micron\ map since most sources have
the highest signal-to-noise in this map.  The source position is
determined by fitting a Gaussian to the flux densities in the pixels
surrounding the pixel containing the peak emission.  As an initial
estimate of the flux density of the source in each SPIRE band, {\tt MADX}
takes the flux density in the pixel closest to the 250-$\mu$m
position.

The high source density on the SPIRE maps means that these flux
estimates often contain contributions from neighbouring sources.  To
mitigate this effect, {\tt MADX} uses the following procedure.  In
each band, {\tt MADX} sorts the sources in order of decreasing flux
density.  The flux density of the brightest source is then more
precisely estimated using the value of the filtered map interpolated
to the exact (sub-pixel) position from the 250-\micron\ map.  Using
this flux estimate, a point-source profile is then subtracted from the
map at this position. Since the bright source is now removed from the
map, any fainter sources nearby should have fluxes that are not
contaminated by the brighter source. The program then moves to the
next brightest source and follows the same set of steps.

The point-source subtraction continues for all sources in sequence,
ordered on the initial flux density estimates. It stops when the PSF
for the faintest source is subtracted. The faintest source considered
is $2.5\sigma$, based on the initial flux and noise estimates.

If two sources with comparable flux are close to each other, then the
algorithm will lead to slightly biased fluxes: the peak of the first
source will include some flux from the wings of the second, and be
overestimated; the psf-subtracted peak of the second source will have
too much subtracted, and so the flux will be slightly
underestimated. The size of these errors is a steep function of the
separation of two sources. For a roughly Gaussian PSF, and two equal
sources separated by twice the FWHM, the first source will have a flux
over-estimated by a factor 1.06 and the second underestimated by
0.997. If they are separated by the FWHM, then the first source is
overestimated by a factor 1.5, and the second is underestimated by a
factor 0.75. At such a small separation, the images are strongly
blended, and so a more sophisticated de-blending algorithm would be
required to improve the flux estimates. For our maps the instrumental
noise is comparable to the confusion noise, and so there is only a
small potential gain from reducing this source of confusion noise. The
error analysis presented in V16 is based on simulated catalogues that
use the de-blending as described above, and so the quoted errors include the
average de-blending errors.

One consequence of these steps is that some sources will have final
250-$\mu$m flux densities less than the original $2.5\sigma$ cut.
Also, most sources are brighter at 250$\mu$m than at the two longer
wavelengths, so the estimates of the flux densities in the
350-$\micron$\ and 500-$\micron$\ bands are typically significantly
lower in signal to noise, and can be negative. Note that a negative
flux measurement is perfectly reasonable so long as the associated
error is comparable.

The released catalogue contains only sources detected at more than
$4\sigma$ significance in any of the bands.  At $4\sigma$, we feel
confident that every catalogue entry corresponds to a real
astrophysical source.  We present flux measurements for all of the
bands for these sources, even if the measurements are negative. We
retain these negative measurements so that the distribution of fluxes
in the catalogues is consistent with the errors, and not truncated at
an arbitrary limit; we do not report `upper limits'.

\begin{figure*} 
   \includegraphics[width=\textwidth, trim=0mm 5mm 0mm 20mm]{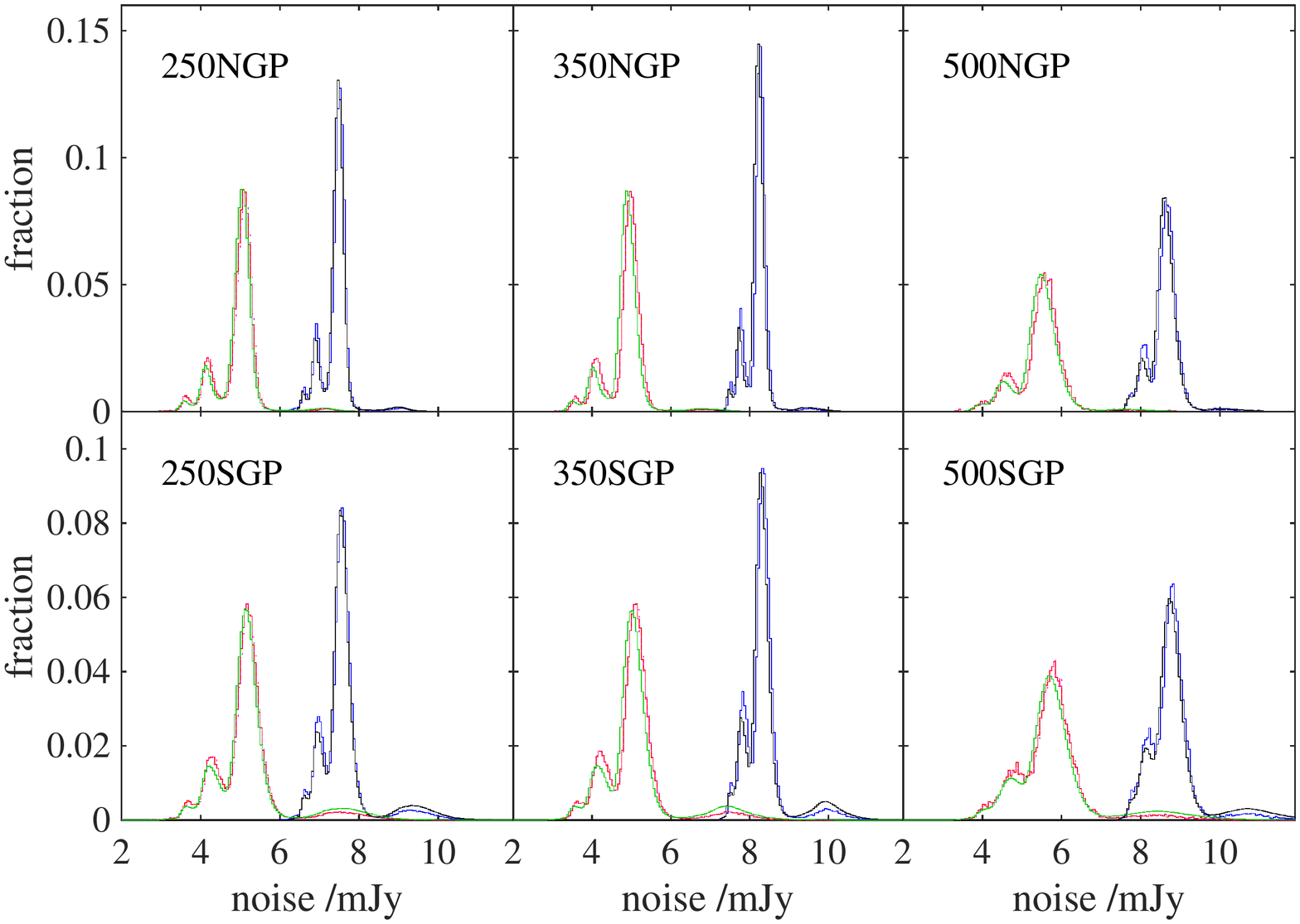}
   \caption{\protect\label{fig_noise} The distribution of instrumental
     and total noise for the 250-\micron, 350-\micron\ and
     500-\micron\ bands for the NGP and SGP fields.  Green shows the
     instrumental noise and black the total noise for all pixels; red
     shows the instrumental noise and blue the total noise at the
     positions of all sources. The multiple peaks are the results of
     our tiling strategy. The main peak corresponds to the large
     fraction of the survey area that was covered by two individual
     {\it Herschel} observations (S17). The smaller peaks correspond
     to the small fraction of the survey area that was either covered
     by more than two observations or, in the case of one end of the
     SGP (S17), a single observation (the small peak at the right in
     the bottom panels.}
\end{figure*}

\begin{figure} 
  \includegraphics[width=0.6\textwidth,trim=25mm 15mm -10mm 25mm]{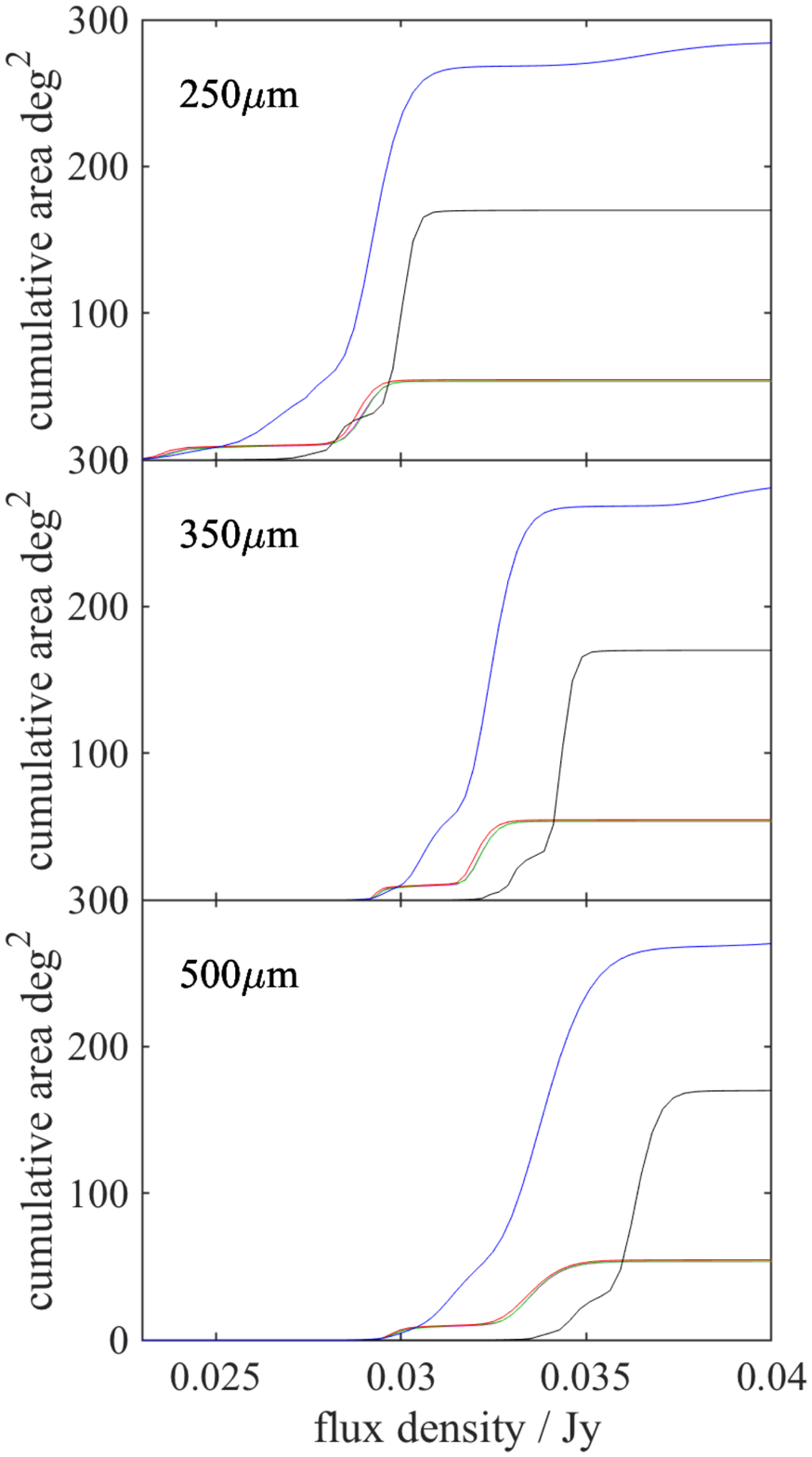}
  \caption{\protect\label{fig_areas} The relationship between area and
    4$\sigma$ flux-density limit for the H-ATLAS fields: NGP - black;
    SGP - blue; GAMA9 - magenta; GAMA12 - green and GAMA15 - cyan.
    The more sensitive areas correspond to the tile overlaps in each
    field.  The westerly end of SGP has only a single SPIRE
    observation, which explains the kink at high flux densities in the
    blue line in these panels.}
\end{figure} 

\section{Photometry}

\subsection{Point sources}

\subsubsection{SPIRE}

V16 carried out extensive simulations to determine the errors on flux
density estimates for point sources in the GAMA fields.  The data for
the NGP and SGP fields were taken in the same mode with the same
observing strategy as the data for the GAMA fields presented by
V16. This means that the statistical properties of the data are
essentially identical to the V16 maps, and so we can directly apply
the V16 results to our current data.  The only potential difference is
that the current maps have some areas where the total coverage has
more than four observations, which was the maximum coverage in V16. As
shown in Table~\ref{tab_areas}, the area with $N_{\mathrm{scan}}>4$
corresponds to less than 0.5\% of the total, and so makes a negligible
difference to the overall statistics of the catalogue.

V16 followed the simple procedure of injecting artificial sources of
known flux density into the real maps and then using {\tt MADX} to
estimate their flux densities (Section 2.2). They found that at
250\mic, the detection wavelength, the confusion noise varies as a
function of source flux density and gave a simple formula to
approximate this: \smallskip
\begin{equation}
\sigma_{\mathrm{con}250} = \sqrt{\min(0.0049,f_{250}/5.6)^2 +
  0.00253^2} \ \ \mathrm{Jy}.
\end{equation}
\smallskip
\noindent They found that at 350\mic\ and 500\mic\ the confusion noise
is roughly constant, with $\sigma_{\mathrm{con}350} = 0.00659$~Jy and
$\sigma_{\mathrm{con}500} = 0.00662$~Jy.

We combine the instrumental and confusion noise to estimate the flux
uncertainty for each individual source: we used these formulae to
estimate the confusion noise at the flux level of the source; we used
the maps of the instrumental noise (Section 2.2) to estimate the
instrumental noise at the position of the source; and then added the
confusion and instrumental noise in quadrature to give the total flux
uncertainty for the source.

Our strategy of creating the H-ATLAS survey from overlapping tiles
(S17) means that the instrumental noise varies systematically between
different areas of the maps.  Fig.~\ref{fig_noise} shows histograms
of instrumental noise and total noise (instrumental noise plus
confusion noise) for all pixels and at the positions of all sources.
The multiple peaks are the results of our tiling strategy. The main
peak corresponds to the large fraction of the survey area that was
covered by two individual {\it Herschel} observations (S17). The
smaller peaks at lower noise correspond to the smaller fraction of the
survey area that was covered by more than two observations. The small
peak at higher noise in the SGP field corresponds to the area at the
western end that was covered by a single observation (S17).

The variation of noise across the maps means that the 4$\sigma$
flux-density limit varies over the fields, and hence the available
area depends on the chosen flux density limit. Fig.~\ref{fig_areas}
shows the relationship between area and flux-density limit for each of
the H-ATLAS fields, including the GAMA fields.

\subsubsection{PACS}

As in V16, we used aperture photometry to estimate the flux densities
in the two PACS bands. We did this for two reasons. First, the PACS
PSF for our observing mode (fast-parallel scan mode) is not well
determined near its peak (see V16 and S17 for extensive
discussions). Second, if we estimated the 100- and 160-$\micron$ flux
densities at the 250-$\mu$m position, as we did for the 350- and
500-$\mu$m bands, we would be likely to significantly underestimate
the flux density because of the higher resolution of the PACS maps.

V16 describes an extensive investigation of the optimum aperture size,
and we follow that paper in using an aperture with a radius equal to
the FWHM, which is 11.4 arcsec for 100\mic\ and 13.7 arcsec
for 160\mic. Although the ``sky'' level has already been subtracted with
{\tt nebuliser}, we subtracted the mean value from each image before
carrying out the photometry,
to ensure that the statistical properties of the sources in the catalogues
are not affected by any residual errors in the sky subtraction.
To
provide an accurate treatment of the contribution from fractional
pixels near aperture boundaries, we divided each pixel into 16, and
assigned one sixteenth of the flux density in each sub-pixel,
corresponding to a nearest-pixel interpolation. Then the flux density
from each sub-pixel that lies within the aperture is added together to
produce the total aperture flux.  We also tried bilinear, and bicubic
interpolation methods and found negligible differences in the
resulting aperture fluxes.  Since only $\simeq$10\% of the SPIRE
sources were clearly detected on the PACS images, we centred the
aperture on the 250-$\mu$m position.

We corrected the aperture flux densities to total flux densities using
the table of the encircled energy fraction (EEF) described in V16 and
available at \url{http://www.h-atlas.org/}.  We made a further
correction to allow for the effect of the errors on the 250-$\micron$
positions, since any error in the position will lead to the small PACS
apertures missing flux. V16 describes simulations of this effect, and
we follow that paper in compensating for this effect by multiplying
the flux densities by 1.1 and 1.05 at 100 and 160 $\mu$m,
respectively.

We describe how we estimated the errors on these flux estimates in the
following subsection.

\subsection{Extended sources} 

The approach in Section 3.1 gives optimal flux density estimates for
point sources, but will substantially underestimate the flux density
of extended sources.  As in V16, we used the $r-$band sizes of optical
counterparts to the {\it Herschel} sources to indicate which sources
are likely to require aperture photometry rather than the methods
described in the last section.  We followed different methods for the
NGP and the SGP because of the lack of a comprehensive identification
analysis for the SGP. We estimate aperture photometry for extended
sources in both PACS and SPIRE bands.

\subsubsection{The NGP}

In the NGP, F17 carried out a search for optical counterparts to the
{\it Herschel} sources on the $r$-band images of the Sloan Digital Sky
Survey (SDSS) which was almost exactly the same as that carried out by
Bourne et al. (2016) for the H-ATLAS GAMA fields.  Our initial list of
NGP sources that might require aperture photometry were the sources
with optical identifications with reliability $R>0.8$ from F17.

In our previous data release (V16) we calculated the sizes of our
apertures from the SDSS parameter $\mathtt{isoA\_r}$, which was
available in SDSS DR7. However, this parameter was not available in
SDSS DR10, on which F17 based their analysis.  After an investigation
of the various size measurements available in DR10, we found that the
parameter $\mathtt{petroR90\_r}$, the 90\% Petrosian radius, met our
needs since there is a simple scaling between it and
$\mathtt{isoA\_r}$, with
$\mathtt{isoA\_r} \simeq 1.156 \, \mathtt{petroR90\_r}$.
The scale-factor 1.156 is derived from a simple fit to
$\mathtt{isoA\_r}$ as a function of $\mathtt{petroR90\_r}$.

We considered that for H-ATLAS sources with optical counterparts with
$\mathtt{petroR90\_r}$ less than 8.6 arcsec (equivalent to the value
of $\mathtt{isoA\_r}$ of 10 arcsec used in V16), the source is still
unlikely to be extended in the SPIRE bands, and for these H-ATLAS
sources we preferred to adopt the flux densities in the SPIRE bands produced by
{\tt MADX} (Section 3.1.1).  However, if the H-ATLAS source had an
optical counterpart with $\mathtt{petroR90\_r}$ greater than 8.6
arcsec, we measured aperture photometry for the SPIRE bands. We
calculated the radius of the aperture using the same formula as V16
(with $\mathtt{isoA\_r}$ replaced by $\mathtt{petroR90\_r}$):
\smallskip
\begin{equation} 
r_\mathrm{ap} = \sqrt{ \mathtt{FWHM}^2 + {(1.156
    \ \mathtt{petroR90\_r})}^2}\ , 
\end{equation}
where $\mathtt{FWHM}$ is the full-width at half-maximum of the
point-spread function for the passband being measured, and all radii
are measured in arcsec. As discussed
above (Section 3.1.2), we also use aperture photometry in the PACS
bands for sources without reliable optical counterparts, using an
aperture with a radius equal to the FWHM.

After calculating the aperture using equation 2, we visually compared
it with the 250-$\mu$m emission from the source, since in some case
the aperture is not well-matched to the 250-$\mu$m emission, either
being too small, too large, with the wrong shape or including the flux
from a neighbouring galaxy (see V16 for examples).  In these cases, we
chose a more appropriate aperture for the galaxy, which may involve
changing the radius or changing to an elliptical aperture.  We also
visually inspected the 3000 sources with the brightest 250-$\mu$m flux
densities from {\tt MADX} in order to check whether there were any
obvious additional extended sources.  For these sources too, we chose
appropriate apertures to include all of the emission.  In total, for
the NGP there are 77 of these ``customised apertures''.  The
semi-major, semi-minor axes and position angles of these customised
apertures are given as part of the data release.

We centred the apertures on the optical positions, since these are
more accurately determined than the {\it Herschel} positions.  Although the
``sky'' level on both the PACS and SPIRE images has already been
subtracted with {\tt nebuliser}, 
we subtracted the mean value
from each image before carrying out the photometry, in order to
avoid residual errors in the sky subtraction affecting the
statistical properties of the catalogues.
As described in Section 3.1.2, we divided each
pixel into 16, assigning one sixteenth of the flux density in each
sub-pixel, and added up the flux density in each sub-pixel within the
aperture.  We corrected the PACS flux densities to total flux
densities using the table EEF described in V16 and available at
\url{http://www.h-atlas.org/}.  We corrected all the SPIRE aperture
flux densities for the fraction of the PSF outside the aperture using
a table of corrections determined from the best estimate of the SPIRE
PSF (Griffin et al. 2013, Valtchanov 2017), which is provided as part
of the data release (see V16 for more details).

We calculated errors in the aperture flux densities from the results
of the Monte Carlo simulation of S17. S17 placed apertures randomly on
the SGP and NGP maps in areas that are made from two individual
observations ($N_{\mathrm{scan}}=2$), varying the aperture radii from
approximately the beam size up to 100 arcsec in 2 arcsec
intervals and using 3000 random positions for each aperture radius.
They found that the error, $\sigma_{\mathrm{ap}}$ in mJy, depends on
the radius the aperture as a double power-law:

\begin{equation}
  \sigma_{\mathrm{ap}}(\mathrm{mJy}) =
  \begin{cases}
      Ar^\alpha , &   \mathrm{if\ } r\le 50'' ,\\
      B(r-50)^\beta + A 50^\alpha , & \mathrm{for\ } r>50''.
    \end{cases}
\end{equation}

The constants $A$, $B$, $\alpha$, and $\beta$ are given in Table~3 of
S17.  We used this equation for the sources on parts of the images
made from two observations. In parts of the images made from more than
two observations the instrumental noise is less; for sources in these
more sensitive parts of the images we used the extensions of equation
3 derived by S17; i.e. equation 4 in S17 for SPIRE and equation 6 in
S17 for PACS. Note that this procedure for estimating uncertainties
intrinsically includes confusion noise and any correlated errors in
the map data.

Finally, we only used the aperture flux density if
it is significantly larger than the point-source estimate, i.e.
\begin{equation}
F_\mathrm{ap}- F_\mathrm{ps}>\sqrt{\sigma_\mathrm{ap}^2-\sigma_\mathrm{ps}^2}
\ .
\end{equation}

In summary, of the 118,986 sources in the NGP, we measured aperture
flux densities at 250\mic\ for 889 sources. 

\subsubsection{The SGP}

For the SGP area no SDSS data exist and we have not carried out the
comprehensive identification analysis that we performed for the other
four fields. Instead, we have carried out a rudimentary identification
analysis using the 2MASS survey (Skrutskie et al. 2006). We first
found a 2MASS galaxy parameter that provides a useful estimate of the
size of the galaxy. We found that the 2MASS parameter ``super-coadd
3-$\sigma$ isophotal semi-major axis'', $\mathtt{sup\_r\_3sig}$, has a
simple scaling with the $\mathtt {isoA\_r}$:
$\mathtt{isoA\_r} \simeq 1.96\, \mathtt{sup\_r\_3sig}$.  The
scale-factor 1.96 is derived from a simple fit to $\mathtt{isoA\_r}$
as a function of $\mathtt{sup\_r\_3sig}$.
  
We found all 2MASS galaxies in the SGP region with
$\mathtt{sup\_r\_3sig}>5.1$ arcsec, equivalent to
$\mathtt{isoA\_r}=10\ $arcsec. There are 6249 of these galaxies. We
then found all H-ATLAS sources in the SGP within 5 arcsec of a 2MASS
galaxy. There are 3444 of these sources. We used the surface-density
of Herschel sources to estimate the probability of a Herschel source
falling within 5 arcsec of a 2MASS galaxy by chance; we estimate that
only 23 (0.7\%) of these matches should not be physical associations
of the H-ATLAS source and the 2MASS galaxy.

For these sources, we calculated the radius of the aperture
to use for photometry using the relationship:
\smallskip
\begin{equation} 
r_\mathrm{ap} = \sqrt{ \mathtt{FWHM}^2 + {(1.96
    \ \mathtt{sup\_r\_3sig})}^2}\ .
\end{equation}
This is the same as equation 2, except for the change
in the parameter used to estimate the size of the galaxy.
In principle we could use $\mathtt{sup\_r\_3sig}$ as our radius
measure for  the sources in the NGP, but SDSS is significantly deeper than 2MASS
and so the measurements are likely to have smaller uncertainties. 

As for the NGP, we then visually compared the apertures with the
250-$\mu$m emission from the source, modifying the aperture when
necessary (see above).  We also visually inspected the 5000 sources
with the brightest 250-$\mu$m flux densities from {\tt MADX} in order to
check whether there were any obvious additional extended sources.  For
these sources, we also chose appropriate apertures to include all of
the emission.  In total, for the SGP there are 142 customised
apertures, for which the details are given as part of the data
release.

In the case of the SGP, we centred the apertures on the 250-$\mu$m
positions rather than on the optical positions. Otherwise we followed
exactly the same procedures to estimate the fluxes and errors as for
the NGP, described in Section 3.2.1.  In summary, of the 118,986
sources in the SGP, we measured aperture flux densities at 250 $\mu$m
for 1452 sources.

\startlongtable
\begin{deluxetable*}{lllll}
  \tablecaption{\protect\label{tab_planck} Comparison of HATLAS with
    \textit{Planck} flux densities (Jy, rounded to 1\,mJy) at 350 and
    $500\,\mu$m.  We have adopted the \textit{Planck} APERFLUX
    photometry as recommended by Planck Collaboration XXVI (2016) for
    these wavelengths. \textit{Planck} 545\,GHz ($550\,\mu$m) flux
    densities, and their errors, have been scaled up by a factor of
    1.35 to convert them to $500\,\mu$m.}  \tabletypesize{\scriptsize}
  \tablewidth{1.4\textwidth} \tablehead{ \colhead{HATLAS\_IAU\_ID} &
    \colhead{F350BEST} & \colhead{F$_{\rm Planck350}$}& \colhead{
      F500BEST} & \colhead{F$_{\rm Planck500}$} }
  \startdata
  HATLASJ125026.0+252947  & $12.128\pm 2.696$  & $12.128\pm 2.696$   & $5.208 \pm 1.160$ & $5.666\pm 0.393$ \\
  HATLASJ125440.7+285619  & $4.974 \pm 0.285$  & $5.254\pm  0.204$   & $1.678 \pm 0.126$ & $1.836\pm 0.176$ \\
  HATLASJ131136.9+225454  & $3.469 \pm 0.304$  & $3.684\pm  0.265$   & $1.279 \pm 0.134$ & $1.328\pm 0.161$ \\
  HATLASJ132035.3+340824  & $2.255 \pm 0.009$  & $2.199\pm  0.270$   & $0.715 \pm 0.009$ & $0.869\pm 0.200$ \\
  HATLASJ133955.6+282402  & $1.701 \pm 0.131$  & $1.563\pm  0.181$   & $0.570 \pm 0.061$ & $0.471\pm 0.130$ \\
  HATLASJ125144.9+254615  & $1.589 \pm 0.215$  & $1.740\pm  0.154$   & $0.639 \pm 0.096$ & $0.842\pm 0.176$ \\
  HATLASJ131503.5+243709  & $1.588 \pm 0.008$  & $2.101\pm  0.298$   & $0.544 \pm 0.009$ & $0.976\pm 0.207$ \\
  HATLASJ133457.2+340238  & $1.311 \pm 0.108$  & $1.454\pm  0.271$   & $0.444 \pm 0.051$ & $0.424\pm 0.275$ \\
  HATLASJ125253.6+282216  & $1.168 \pm 0.099$  & $0.985\pm  0.142$   & $0.359 \pm 0.008$ & $0.512\pm 0.157$ \\
  HATLASJ132815.2+320157  & $1.043 \pm 0.093$  & $1.022\pm  0.280$   & $0.362 \pm 0.044$ &       ---        \\
  HATLASJ134308.8+302016  & $1.032 \pm 0.114$  & $0.605\pm  0.288$   & $0.319 \pm 0.009$ &       ---        \\
  HATLASJ131206.6+240543  & $1.019 \pm 0.029$  & $1.444\pm  0.278$   & $0.385 \pm 0.020$ &       ---        \\
  HATLASJ132255.7+265857  & $0.987 \pm 0.094$  & $0.958\pm  0.228$   & $0.330 \pm 0.045$ &       ---        \\
  HATLASJ131612.2+305702  & $0.925 \pm 0.117$  & $1.498\pm  0.310$   & $0.346 \pm 0.055$ &       ---        \\
  HATLASJ130547.6+274405  & $0.922 \pm 0.118$  & $1.278\pm  0.442$   & $0.363 \pm 0.055$ & $0.760\pm 0.292$ \\
  HATLASJ130514.1+315959  & $0.832 \pm 0.104$  & $0.840\pm  0.165$   & $0.258 \pm 0.008$ &       ---        \\
  HATLASJ130056.1+274727  & $0.769 \pm 0.023$  & $0.934\pm  0.269$   & $0.268 \pm 0.018$ &       ---        \\
  HATLASJ133026.1+313707  & $0.737 \pm 0.100$  & $0.861\pm  0.220$   & $0.267 \pm 0.047$ &       ---        \\
  HATLASJ124610.1+304355  & $0.718 \pm 0.047$  & $0.525\pm  0.342$   & $0.241 \pm 0.009$ &       ---        \\
  HATLASJ130125.2+291849  & $0.701 \pm 0.008$  & $0.854\pm  0.233$   & $0.232 \pm 0.009$ &       ---        \\
  HATLASJ130947.5+285424  & $0.680 \pm 0.122$  & $0.971\pm  0.168$   & $0.220 \pm 0.057$ &       ---        \\
  HATLASJ130617.2+290346  & $0.675 \pm 0.103$  & $1.057\pm  0.288$   & $0.247 \pm 0.049$ &       ---        \\
  HATLASJ131241.9+224950  & $0.650 \pm 0.154$  & $0.230\pm  0.209$   & $0.253 \pm 0.072$ &       ---        \\
  HATLASJ131101.7+293442  & $0.628 \pm 0.089$  & $1.114\pm  0.409$   & $0.189 \pm 0.008$ &       ---        \\
  HATLASJ125108.4+284705  & $0.611 \pm 0.090$  & $0.661\pm  0.285$   & $0.245 \pm 0.043$ &       ---        \\
  HATLASJ133550.1+345957  & $0.602 \pm 0.025$  & $1.056\pm  0.236$   & $0.200 \pm 0.019$ &       ---        \\
  HATLASJ132948.2+310748  & $0.559 \pm 0.017$  & $0.580\pm  0.308$   & $0.190 \pm 0.014$ &       ---        \\
  HATLASJ131730.6+310533  & $0.548 \pm 0.021$  & $0.610\pm  0.253$   & $0.196 \pm 0.016$ &       ---        \\
  HATLASJ131327.0+274807  & $0.520 \pm 0.114$  & $1.149\pm  0.379$   & $0.195 \pm 0.053$ &       ---        \\
  HATLASJ133554.6+353511  & $0.510 \pm 0.079$  & $0.925\pm  0.148$   & $0.187 \pm 0.038$ &       ---        \\
  HATLASJ125008.7+330933  & $0.509 \pm 0.022$  & $0.521\pm  0.209$   & $0.190 \pm 0.017$ &       ---        \\
  HATLASJ131745.2+273411  & $0.500 \pm 0.019$  & $1.178\pm  0.230$   & $0.169 \pm 0.015$ &       ---        \\
	\hline
  HATLASJ235749.9$-$323526 & $24.881\pm 1.894$ & $24.513\pm 0.723$ & $10.667\pm 0.821$ & $11.743\pm 0.566$\\
  HATLASJ003024.0$-$331419 & $22.390\pm 1.140$ & $23.014\pm 0.447$ & $8.103 \pm 0.497$ & $8.624 \pm 0.243$\\
  HATLASJ013418.2$-$292506 & $17.557\pm 1.040$ & $16.747\pm 3.56$  & $5.931 \pm 0.452$ & $5.979 \pm 0.202$\\
  HATLASJ005242.2$-$311222 & $4.922 \pm 0.445$ & $6.425 \pm 0.225$ & $1.788 \pm 0.198$ & $2.892 \pm 0.220$\\
  HATLASJ003415.3$-$274812 & $4.063 \pm 0.407$ & $4.183 \pm 0.341$ & $1.482 \pm 0.180$ & $1.705 \pm 0.181$\\
  HATLASJ234751.7$-$303118 & $3.420 \pm 0.433$ & $3.317 \pm 0.170$ & $1.288 \pm 0.192$ & $1.386 \pm 0.212$\\
  HATLASJ225801.7$-$334432 & $3.294 \pm 0.225$ & $3.627 \pm 0.266$ & $1.149 \pm 0.102$ & $1.404 \pm 0.173$\\
  HATLASJ003658.8$-$292839 & $2.246 \pm 0.009$ & $2.486 \pm 0.382$ & $0.741 \pm 0.009$ & $0.601 \pm 0.227$\\
  HATLASJ224218.1$-$300333 & $1.963 \pm 0.312$ & $1.768 \pm 0.184$ & $0.869 \pm 0.140$ & $0.629 \pm 0.217$\\
  HATLASJ011407.0$-$323908 & $1.622 \pm 0.123$ & $2.097 \pm 0.507$ & $0.629 \pm 0.058$ & $0.763 \pm 0.327$\\
  HATLASJ000833.7$-$335147 & $1.533 \pm 0.235$ & $1.322 \pm 0.309$ & $0.504 \pm 0.107$ & --- \\
  HATLASJ222421.6$-$334139 & $1.519 \pm 0.043$ & $1.819 \pm 0.156$ & $0.481 \pm 0.031$ & $0.953 \pm 0.151$\\
  HATLASJ013906.2$-$295457 & $1.445 \pm 0.034$ & $2.064 \pm 0.374$ & $0.545 \pm 0.024$ & $0.803 \pm 0.231$\\
  HATLASJ011035.6$-$301316 & $1.314 \pm 0.035$ & $1.482 \pm 0.180$ & $0.497 \pm 0.025$ & $0.536 \pm 0.216$\\
  HATLASJ222521.1$-$312116 & $1.251 \pm 0.120$ & $1.551 \pm 0.294$ & $0.480 \pm 0.058$ & --- \\
  HATLASJ014021.4$-$285445 & $1.224 \pm 0.093$ & ---               & $0.421 \pm 0.455$ & $0.990 \pm 0.259$\\
  HATLASJ013150.3$-$330710 & $1.192 \pm 0.118$ & $2.073 \pm 0.308$ & $0.397 \pm 0.056$ & $0.983 \pm 0.198$\\
  HATLASJ010612.2$-$301041 & $1.150 \pm 0.131$ & $1.074 \pm 0.294$ & $0.388 \pm 0.062$ & --- \\
  HATLASJ014744.6$-$333607 & $1.089 \pm 0.035$ & $1.060 \pm 0.287$ & $0.365 \pm 0.025$ & --- \\
  HATLASJ005747.0$-$273004 & $1.073 \pm 0.032$ & $2.043 \pm 0.560$ & $0.445 \pm 0.023$ & $1.118 \pm 0.258$\\
  HATLASJ010456.0$-$272545 & $1.035 \pm 0.032$ & $1.364 \pm 0.266$ & $0.365 \pm 0.023$ & --- \\
  HATLASJ225956.7$-$341415 & $1.033 \pm 0.118$ & $1.218 \pm 0.267$ & $0.314 \pm 0.009$ & --- \\
  HATLASJ011101.1$-$302620 & $0.997 \pm 0.033$ & $0.770 \pm 0.182$ & $0.362 \pm 0.024$ & --- \\
  HATLASJ222610.7$-$310840 & $0.956 \pm 0.093$ & $0.621 \pm 0.349$ & $0.342 \pm 0.046$ & --- \\
  HATLASJ011429.7$-$311053 & $0.917 \pm 0.114$ & $1.069 \pm 0.280$ & $0.331 \pm 0.054$ & --- \\
  HATLASJ012658.0$-$323234 & $0.845 \pm 0.097$ & $0.982 \pm 0.341$ & $0.296 \pm 0.046$ & --- \\
  HATLASJ012315.0$-$325028 & $0.806 \pm 0.031$ & $0.744 \pm 0.465$ & $0.277 \pm 0.023$ & --- \\
  HATLASJ002354.3$-$323210 & $0.803 \pm 0.030$ & $1.108 \pm 0.193$ & $0.314 \pm 0.022$ & --- \\
  HATLASJ002938.2$-$331534 & $0.745 \pm 0.111$ & $0.747 \pm 0.363$ & $0.296 \pm 0.052$ & --- \\
  HATLASJ011122.3$-$291404 & $0.727 \pm 0.031$ & $1.547 \pm 0.455$ & $0.278 \pm 0.022$ & --- \\
  HATLASJ005457.3$-$320115 & $0.719 \pm 0.008$ & $0.780 \pm 0.258$ & $0.245 \pm 0.009$ & --- \\
  HATLASJ012434.5$-$331024 & $0.640 \pm 0.030$ & $0.946 \pm 0.288$ & $0.204 \pm 0.022$ & --- \\
  HATLASJ230549.0$-$303642 & $0.637 \pm 0.085$ & $0.868 \pm 0.345$ & $0.191 \pm 0.009$ & --- \\
  HATLASJ000254.5$-$341407 & $0.572 \pm 0.026$ & $1.160 \pm 0.409$ & $0.207 \pm 0.020$ & --- \\
  HATLASJ001112.7$-$333442 & $0.499 \pm 0.036$ & $0.555 \pm 0.221$ & $0.171 \pm 0.026$ & --- \\
  HATLASJ003651.4$-$282200 & $0.466 \pm 0.065$ & $0.438 \pm 0.324$ & $0.162 \pm 0.032$ & --- \\
  HATLASJ010723.3$-$324943 & $0.448 \pm 0.047$ & $0.839 \pm 0.881$ & $0.141 \pm 0.009$ & --- \\
  HATLASJ225739.6$-$293730 & $0.433 \pm 0.009$ & $0.975 \pm 0.293$ & $0.292 \pm 0.009$ & --- \\
  HATLASJ004806.7$-$284818 & $0.407 \pm 0.008$ & $0.069 \pm 0.238$ & $0.126 \pm 0.009$ & --- \\
  HATLASJ235939.7$-$342829 & $0.352 \pm 0.064$ & $0.733 \pm 0.215$ & $0.095 \pm 0.009$ & --- \\
  HATLASJ005852.3$-$281812 & $0.349 \pm 0.019$ & $0.209 \pm 0.490$ & $0.113 \pm 0.015$ & --- \\
  HATLASJ233007.0$-$310738 & $0.213 \pm 0.040$ & $0.433 \pm 0.305$ & $0.101 \pm 0.022$ & --- \\
\enddata
\end{deluxetable*}

\begin{figure*} 
  \includegraphics[width=0.5\textwidth, trim=40mm .0mm 40.0mm .0mm, clip=True]{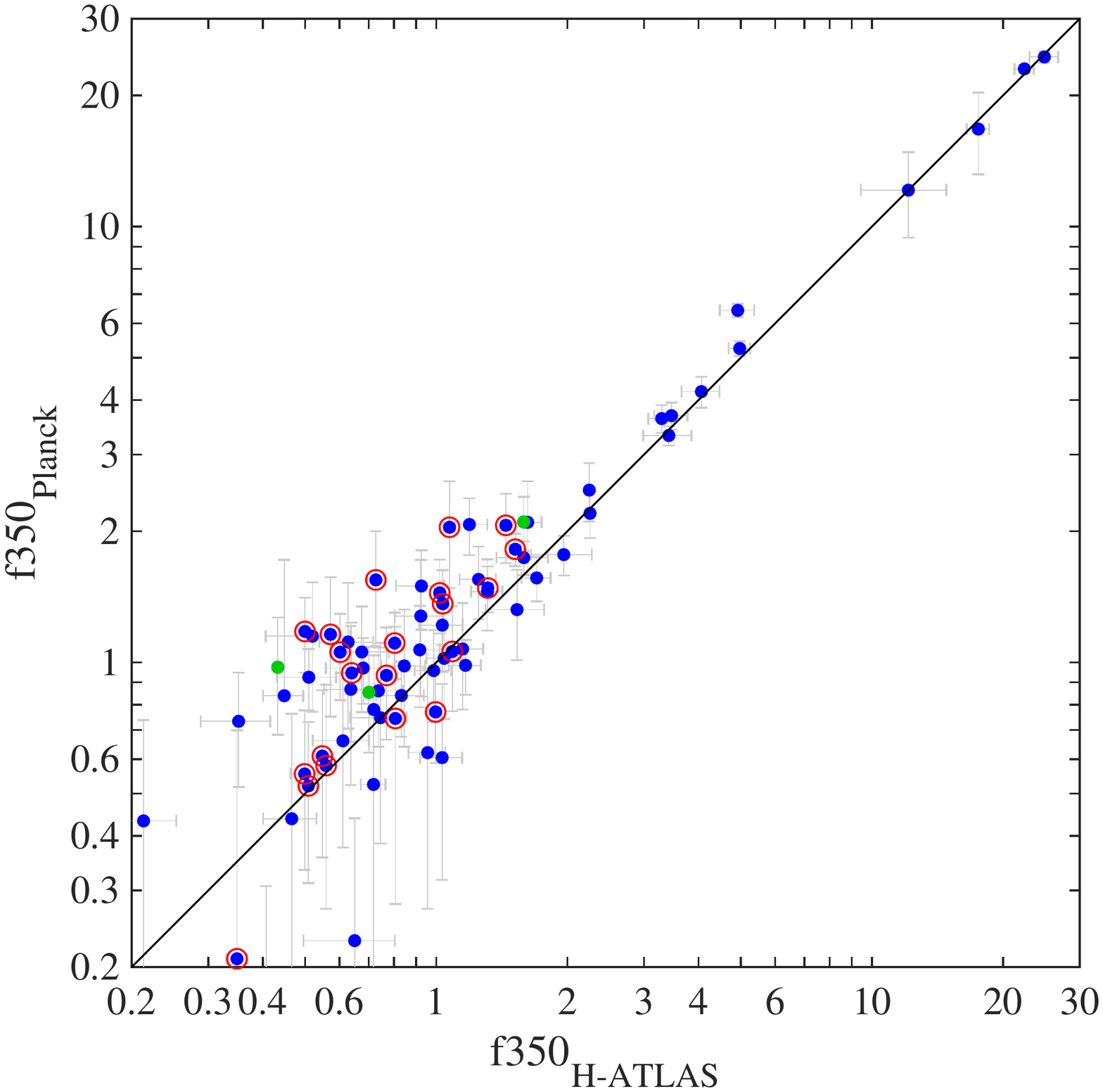}
  \includegraphics[width=0.5\textwidth, trim=40mm .0mm 40.0mm .0mm, clip=True]{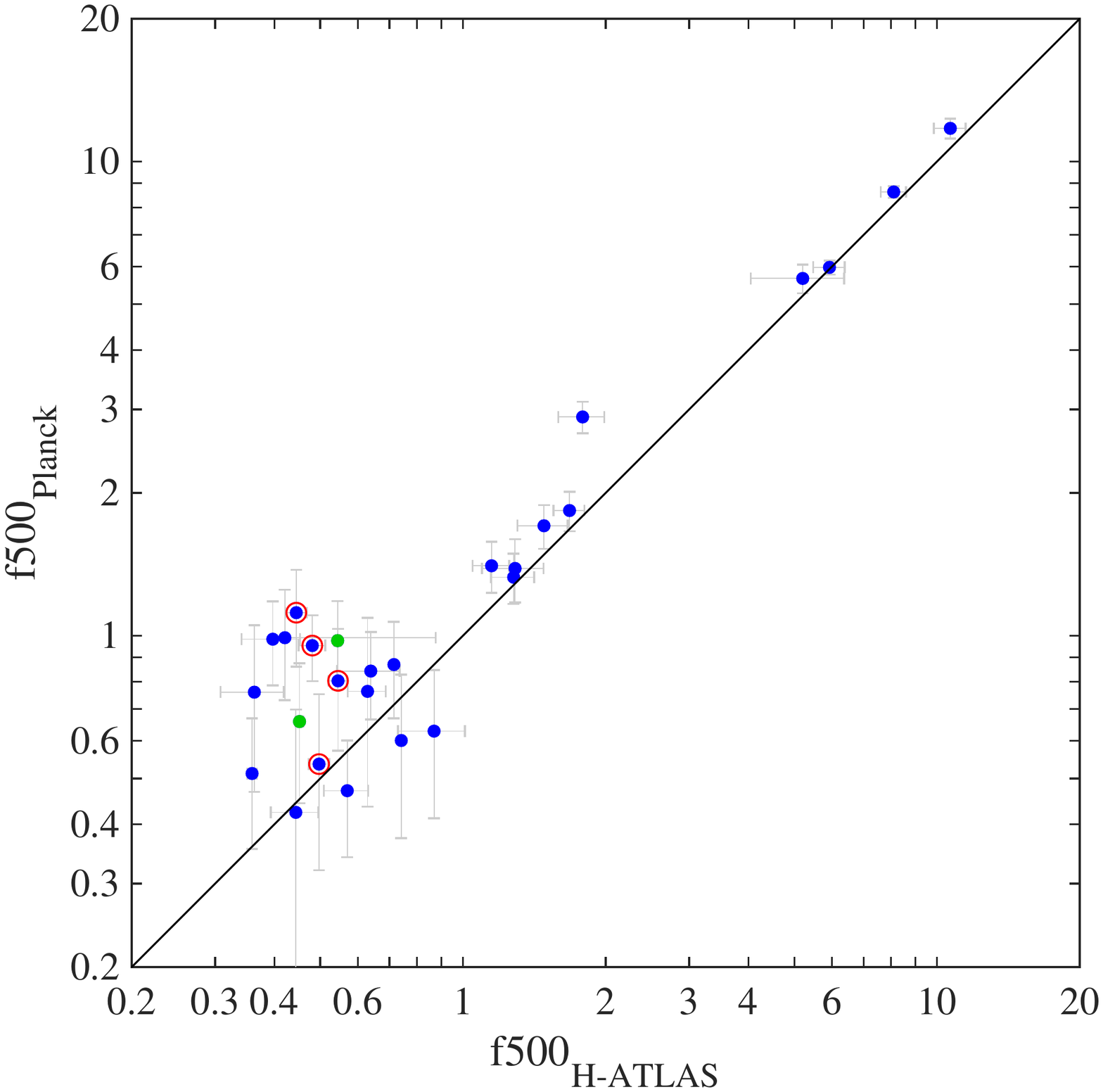}
  \caption{\protect\label{fig_planck} Comparison between H-ATLAS and
    {\it Planck} flux density measurements for 350\mic\ and 500\mic\
    bands. Green points show H-ATLAS point sources; blue points show
    H-ATLAS extended sources; extended sources which have {\it not}
    been given a custom aperture are circled in red. The majority of
    the Planck sources are so large that they have been given custom
    apertures in the H-ATLAS catalogue: 48/73 at 350\mic; and 21/27 at
    500\mic.  }
\end{figure*}

\subsection{ Comparison to Planck photometry}

Estimating the flux density of extended sources is sensitive to the
background subtraction and choice of aperture size, so it is useful to
compare our extended source fluxes to other measurements available. In
particular for the 350\mic\ and 500\mic\ bands, we have compared to
the compact source catalogue from the {\it Planck}\footnote{Planck
  (\url{http://www.esa.int/Planck}) is an ESA science mission with
  instruments and contributions directly funded by ESA Member States,
  NASA, and Canada.}  survey (Planck Collaboration XXVI, 2016).  Given
the low-surface density of sources, a simple positional match is
sufficient to cross-identify sources in common. We find 32 matches in
the NGP, and 42 in the SGP as listed in Table~\ref{tab_planck} and
plotted in Fig.~\ref{fig_planck}. Most sources, 49/74, are large
enough that we had visually inspected them and assigned custom
apertures (see \S 3.2.1 and \S 3.2.2), 21 are extended and have
automatically assigned apertures, and 4 are point sources.  We have
adopted the \textit{Planck} APERFLUX photometry as recommended by
Planck Collaboration XXVI (2016) for these wavelengths. The
\textit{Planck} 545\,GHz ($550\,\mu$m) flux densities, and their
errors, have been scaled up by a factor of 1.35 to convert them to
$500\,\mu$m.

As seen in Fig.~\ref{fig_planck}, there is a very good correspondence
between the measurements with no significant systematic offsets or
non-linearity. The {\it Planck} fluxes do appear to be slightly higher
than the HATLAS fluxes at less than 1Jy, but summing over all sources,
the offset is less than the 2-$sigma$ significance level. It is likely
that this is a result of flux boosting in the {\it Planck} catalogue:
if the Planck fluxes were each shifted lower by half of their quoted
uncertainty, there would be no offset. After shifting to remove the
offset, the scatter between the measurements is consistent with the
quoted uncertainties, with $\chi^2=71$ and 72 degrees of freedom. 


The Planck Collaboration XXVI (2016) quote 90\% completeness limits of
791\,mJy and 555\,mJy for the 350- and 550-\micron\ catalogues
respectively. The comparison with the HATLAS catalogue suggests 90\%
completeness down to {\tt F350BEST}=650\,mJy.  For the Planck
550-\micron\ catalogue, the quoted 90\% completeness limit of
555\,mJy, corresponds to 749\,mJy at 500\mic; the comparison with the
HATLAS catalogue suggests 90\% completeness down to {\tt
  F500BEST}=400\,mJy.  Despite the relatively small number of sources,
our comparison suggests that quoted Planck limits are quite
conservative.

\subsection{Colour corrections and flux calibration}

The large wavelength range within each of the SPIRE pass bands means
that both the size of the PSF and the power detected by SPIRE depend
on the spectral energy distribution of the source.  The SPIRE
data-reduction pipeline and ultimately our flux densities are based on
the assumption that the flux density of a source varies with frequency
as $\nu^{-1}$.  If the user knows the SED of a source, the flux
densities should be corrected using corrections from either table 5.7
or 5.8 from the SPIRE
handbook\footnote{\url{http://herschel.esac.esa.int/Docs/SPIRE/spire\_handbook.pdf}}
(Valtchanov 2017). It is important to apply these corrections, since
they can be quite large: for a point source with a typical dust
spectrum ($T=20$K, $\beta=2$) the multiplicative correction is 0.96,
0.94 and 0.90 at 250, 350 and 500 $\mu$m, respectively. The catalogue
fluxes have had no colour correction applied.

As with SPIRE, the PACS flux densities are also based on the
assumption that flux density of the source is proportional to
$\nu^{-1}$, and a correction is required for sources which follow a
different SED. The required corrections are described in the PACS
Colour-Correction
document\footnote{\url{http://herschel.esac.esa.int/twiki/pub/Public/PacsCalibrationWeb/cc\_report\_v1.pdf}}.

On top of all other errors, there is an additional error due to the
uncertain photometric calibration of {\it Herschel}. As in V16, we assume
conservative calibration errors of 5.5\% for the three SPIRE wavebands
and 7\% for PACS (see V16 for more details).

\section{The Catalogues}

We included all sources in the catalogues that were detected above
4$\sigma$ in one or more of the three SPIRE bands: 250, 350 and 500
$\mu$m. We eliminated all sources from the original list of point
sources produced by {\tt MADX} if they fell within the aperture of an
extended source.  The parameters available for each source are listed
in Table~\ref{tab_catalogue}. If a source is included in the 4$\sigma$
catalogue, flux measurements are presented for all bands, with no
censorship at low signal to noise. This means that some flux
measurements are negative; these are not flagged in any way, but
simply listed with the corresponding uncertainty.

Since the PACS instrument is not exactly aligned with the SPIRE
instrument, there are some sources in the catalogue that have no PACS
coverage; the PACS fluxes for these sources are flagged as $-1$. (No
real sources have measurement $<-0.3$Jy, so there is no possibility of
confusion between the flagged sources and negative flux measurements).

All of the H-ATLAS fields were observed at least twice, making it
possible to search for moving sources such as asteroids. We found nine
asteroids in the GAMA fields (V16), eliminating these from the final
catalogue. We carried out the same search for the NGP and SGP but
found no moving objects. Both the NGP and SGP fields are at much
higher ecliptic latitude than the GAMA fields, so it is perhaps not
surprising that we find no more solar-system objects.

The sources in the final catalogues are almost all extra-galactic
sources. We carried out a search for clusters of sources in all the
H-ATLAS fields (Eales et al. in preparation).  In the GAMA9 field, we
found several groups of sources that are likely to be clusters of
pre-stellar cores, implying that the catalogue for this field is likely
to contain a few tens of Galactic sources.  However, we found no
similar clusters in the other fields, which makes sense, since the
GAMA9 field is at a much lower Galactic latitude than the other
fields. Pre-stellar cores are therefore likely to be a very minor
contaminant to the catalogues for these fields. There are a few debris
disks and AGB stars in the catalogues, and an incomplete list is given
in Table~\ref{tab_stars}.  However, well over 99\% of the sources are
extra-galactic. The extra-galactic sources range from galaxies at
redshift 6 (Fudamoto et al. 2017) to nearby galaxies, such as the
spectacular spiral galaxy, NGC 7793, which is in the centre of the
SGP, and one of the brightest galaxies in the nearby Sculptor group.

\begin{deluxetable*}{ll} {
    \tablecaption{\protect\label{tab_catalogue} Data columns in the
    H-ATLAS catalogue files. For NGP sources associated to SDSS and
    UKIDSS sources, there are further columns listing optical and NIR
    properties, as detailed in F17.}
  \tablecolumns{2} \tablewidth{0pt} \tabletypesize{\scriptsize}
  \tablehead{ \colhead{Column
        Name} & \colhead{Description}} \startdata HATLAS\_IAU\_ID &
    \parbox[t]{13cm}{Source name using the IAU
      standard }\\
    IDNAME & \parbox[t]{13cm}{Internal catalogue name based on the
      source number and field name} \\
    RA & \parbox[t]{13cm}{Right Ascension in degrees based on the
      HATLAS data }\\
    DEC & \parbox[t]{13cm}{Declination in degrees based on the
      HATLAS data}\\
    F250 & \parbox[t]{13cm}{Point source 250 \mic\ flux estimate in
      Jy}\\
    F350 & \parbox[t]{13cm}{Point source 350 \mic\ flux estimate in
      Jy}\\
    F500 & \parbox[t]{13cm}{Point source 500 \mic\ flux estimate in
      Jy}\\
    E250 & \parbox[t]{13cm}{Uncertainty on point source 250 \mic\ flux
      in Jy (includes both confusion and
      instrumental noise)}\\
    E350 & \parbox[t]{13cm}{Uncertainty on point source 350 \mic\ flux
      in Jy (includes both confusion and
      instrumental noise)}\\
    E500 & \parbox[t]{13cm}{Uncertainty on point source 500 \mic\ flux
      in Jy (includes both confusion and
      instrumental noise)}\\
    F250BEST& \parbox[t]{13cm}{Best estimate of 250 \mic\ flux in Jy:
      point source if unresolved; aperture flux if resolved. }\\
    E250BEST& \parbox[t]{13cm}{Uncertainty on best estimate 250 \mic\
      flux in Jy (includes both confusion and
      instrumental noise)}\\
    AP250 & \parbox[t]{13cm}{Semi-major axis of 250 \mic\ band
      aperture in arcsecs. $-99$ if point source flux used. }\\
    F350BEST & \parbox[t]{13cm}{Best estimate of 350 \mic\ flux in Jy:
      point source if unresolved; aperture flux if resolved. }\\
    E350BEST & \parbox[t]{13cm}{Uncertainty on best estimate 350 \mic\
      flux in Jy (includes both confusion and
      instrumental noise)}\\
    AP350 & \parbox[t]{13cm}{Semi-major axis of 350 \mic\ band
      aperture in arcsecs. $-99$ if point source flux used}\\
    F500BEST& \parbox[t]{13cm}{Best estimate of 500 \mic\ flux in Jy:
      point source if unresolved; aperture flux if resolved. }\\
    E500BEST& \parbox[t]{13cm}{Uncertainty on best estimate 500 \mic\
      flux in Jy (includes both confusion and
      instrumental noise)}\\
    AP500 & \parbox[t]{13cm}{Semi-major axis of 350 \mic\ band
      aperture in arcsecs. $-99$ if point source flux used}\\
    F100BEST& \parbox[t]{13cm}{Best estimate of 100 \mic\ flux in
      Jy. The value $-1$ indicates that there is no PACS coverage for the
      source}\\
    E100BEST& \parbox[t]{13cm}{Uncertainty on 100 \mic\ flux in
      Jy. The
      value $-1$ indicates that there is no PACS coverage for the source}\\
    AP100 & \parbox[t]{13cm}{Semi-major axis of aperture used for
      100 \mic\ flux, in arcsecs. }\\
    F160BEST& \parbox[t]{13cm}{Best estimate of 160 \mic\ flux in
      Jy. The value $-1$ indicates that there is no PACS coverage for the
      source}\\
    E160BEST& \parbox[t]{13cm}{Uncertainty on 160 \mic\ flux in
      Jy. The
      value $-1$ indicates that there is no PACS coverage for the source}\\
    AP160 & \parbox[t]{13cm}{Semi-major axis of aperture used for
      160 \mic\ flux, in arcsecs. }\\
    AP\_RMIN &\parbox[t]{13cm}{Semi-minor axis of aperture in
      arcsecs. Set only for custom apertures. The value $-99$ flags that
      either an automatically calculated circular aperture has been
      used, or no aperture has been used.}  \\
    AP\_PA & \parbox[t]{13cm}{Position angle of major axis of aperture
      in degrees anti-clockwise from west. Set only for custom
      apertures. The value $-99$ flags that either an automatically
      calculated circular aperture has been used, or no aperture has
      been
      used}\\
    & \\
    \enddata }
\end{deluxetable*} 

\subsection{Statistics of the catalogues}

The catalogue for the NGP covers 177.1 deg$^2$ and contains 118,980
sources, of which 112,069 were detected at $>4\sigma$ at 250 $\mu$m,
46,876 at $>4\sigma$ at 350 $\mu$m and 10,368 at $>4\sigma$ at 500
$\mu$m. The effective sensitivity of the PACS images was much less,
but the catalogues contain flux-density measurements at 100 $\mu$m and
160 $\mu$m for all the sources in the catalogue, even if the
measurements were negative. 5,036 sources were detected at $>3\sigma$
at 100 $\mu$m and 7,046 sources were detected at $>3\sigma$ at 160
$\mu$m.

\begin{deluxetable}{ll}
\tablecaption{Stars detected in H-ATLAS.  \label{tab_stars}}
\tablecolumns{2}
\tablewidth{0pt} \tabletypesize{\scriptsize}
\tablehead{ \colhead{Name} & \colhead{Position}}
\startdata
EY Hya & 08:46:21.4 $+$01:37:53 \\
IN Hya & 09:20:36.7 $+$00:10:53 \\
NU Com & 13:10:08.5 $+$24:36:02 \\
19 PsA & 22:42:22.3 $-$29:21:43 \\
V PsA & 22:55:19.9  $-$29:36:48 \\
S Scl & 00:15:22.4 $-$32:02:44 \\
XY Scl & 00:06:35.9 $-$32:35:38 \\
eta Scl & 00:27:55.9 $-$33:00:27 \\
Y Sci & 23:09:05.7 $-$30:08:04 \\
HD 119617 & 13:43:35.2 $+$35:20:45 \\
R Sci & 01:26:58.2 $-$32:32:37 \\
Fomalhaut & 22:57:39.2 $-$29:37:22 \\
\enddata
\end{deluxetable}

The catalogue for the SGP covers 303.4 deg$^2$ and contains
193,527 sources, of which 182,282 were detected at $>4\sigma$ at 250
$\mu$m, 74,069 at $>4\sigma$ at 350 $\mu$m and 16,084 at $>4\sigma$ at
500 $\mu$m.  8,598 sources were detected at $>3\sigma$ at 100 $\mu$m
and 11,894 sources were detected at $>3\sigma$ at 160 $\mu$m.

\begin{figure}
  \includegraphics[width=0.5\textwidth,trim=0mm 0mm 0mm 20mm]{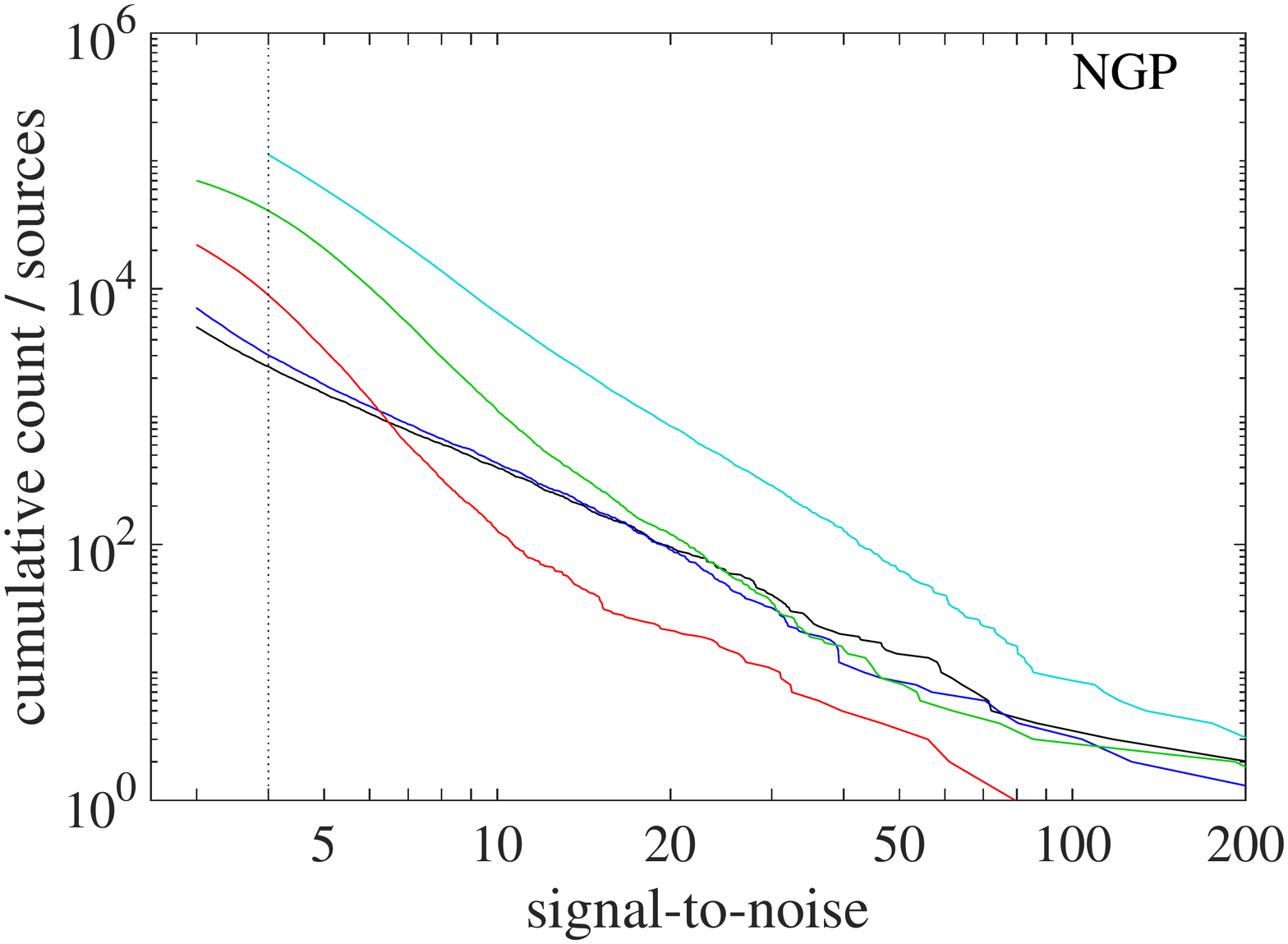}
\includegraphics[width=0.5\textwidth]{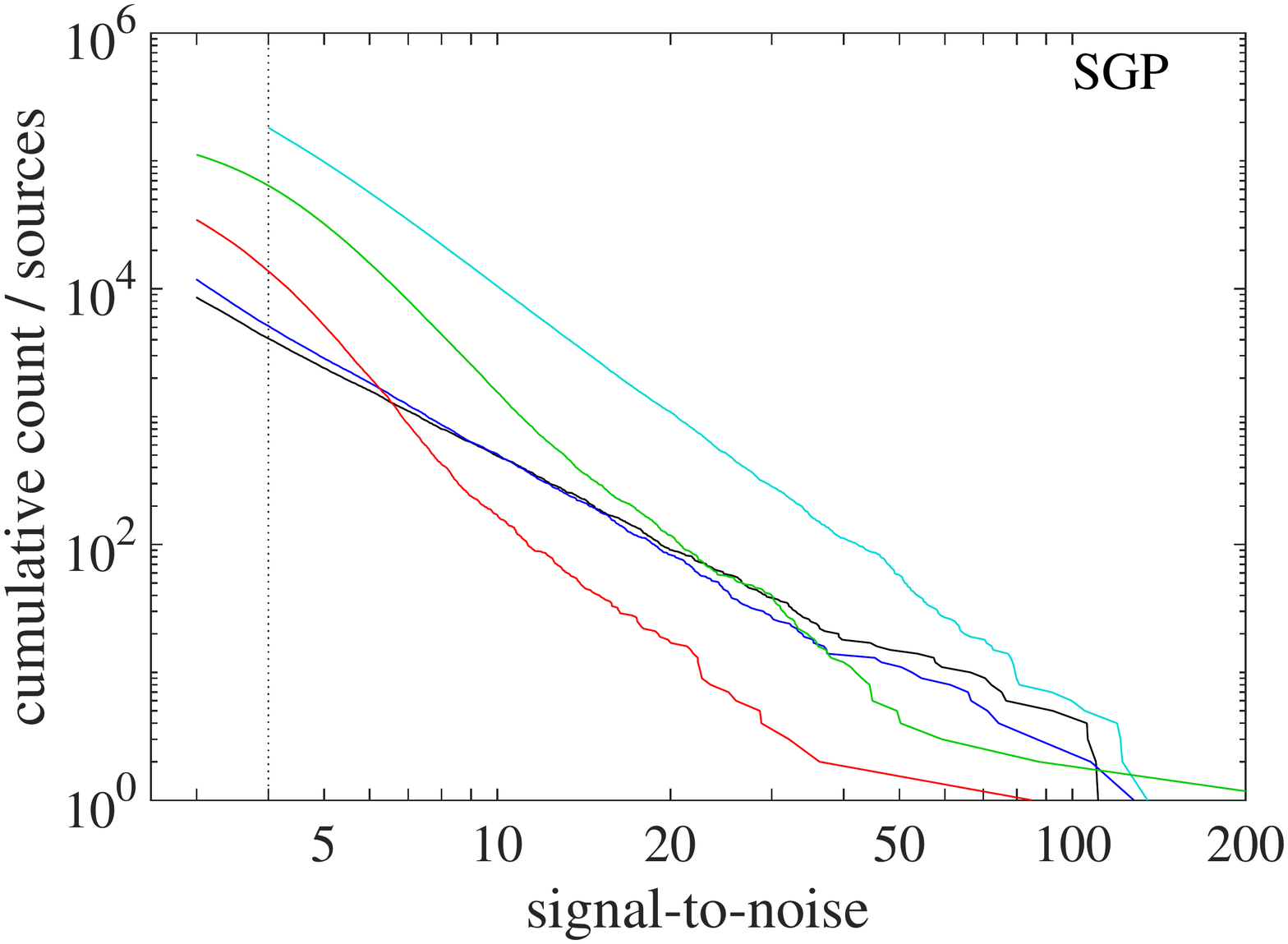}
 \caption{\protect\label{fig_sn} The cumulative number of sources as a function
   of signal-to-noise at 100\mic\ (black), 160\mic\ (blue), 250\mic\ (cyan),
  350\mic\ (green) and 500\mic\ (red). The NGP area is shown in the top panel,
  and the SGP in bottom panel. The vertical dotted line shows the
  4-$\sigma$ limit for the 250\mic\ selection. The other bands are
  truncated at 3-$\sigma$.  
} 
\end{figure}

The cumulative number of sources as a function of signal-to-noise in
the five bands is shown in Fig.~\ref{fig_sn}. The 250-\micron\ band is
the most sensitive, and so has the largest number of detected sources. Of
the PACS bands, the 160-\micron\ band detects more sources above
3-$\sigma$.

\begin{figure}
  \includegraphics[width=0.5\textwidth,trim=0mm 0mm 0mm 20mm]
  {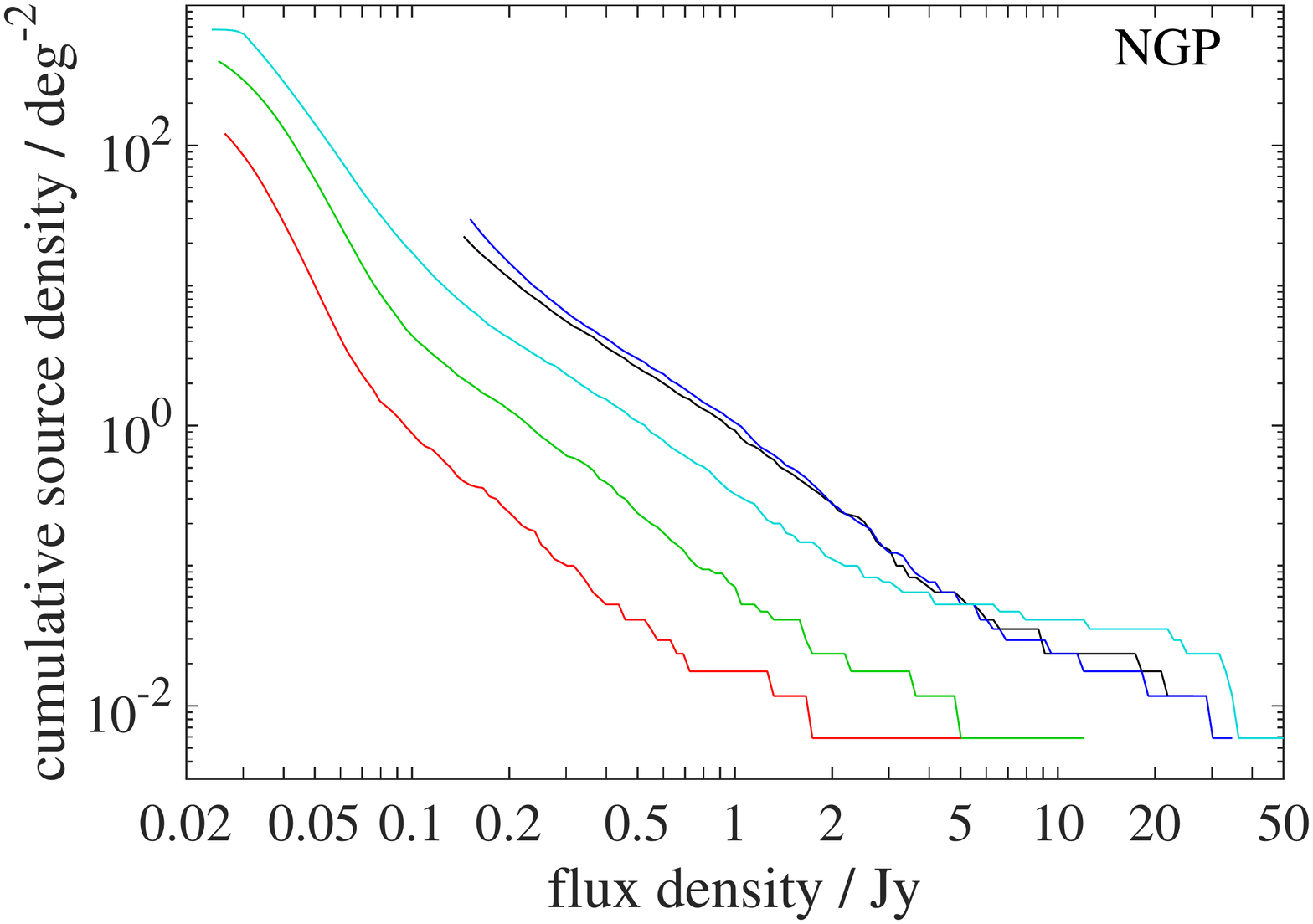}\\
  \includegraphics[width=0.5\textwidth]{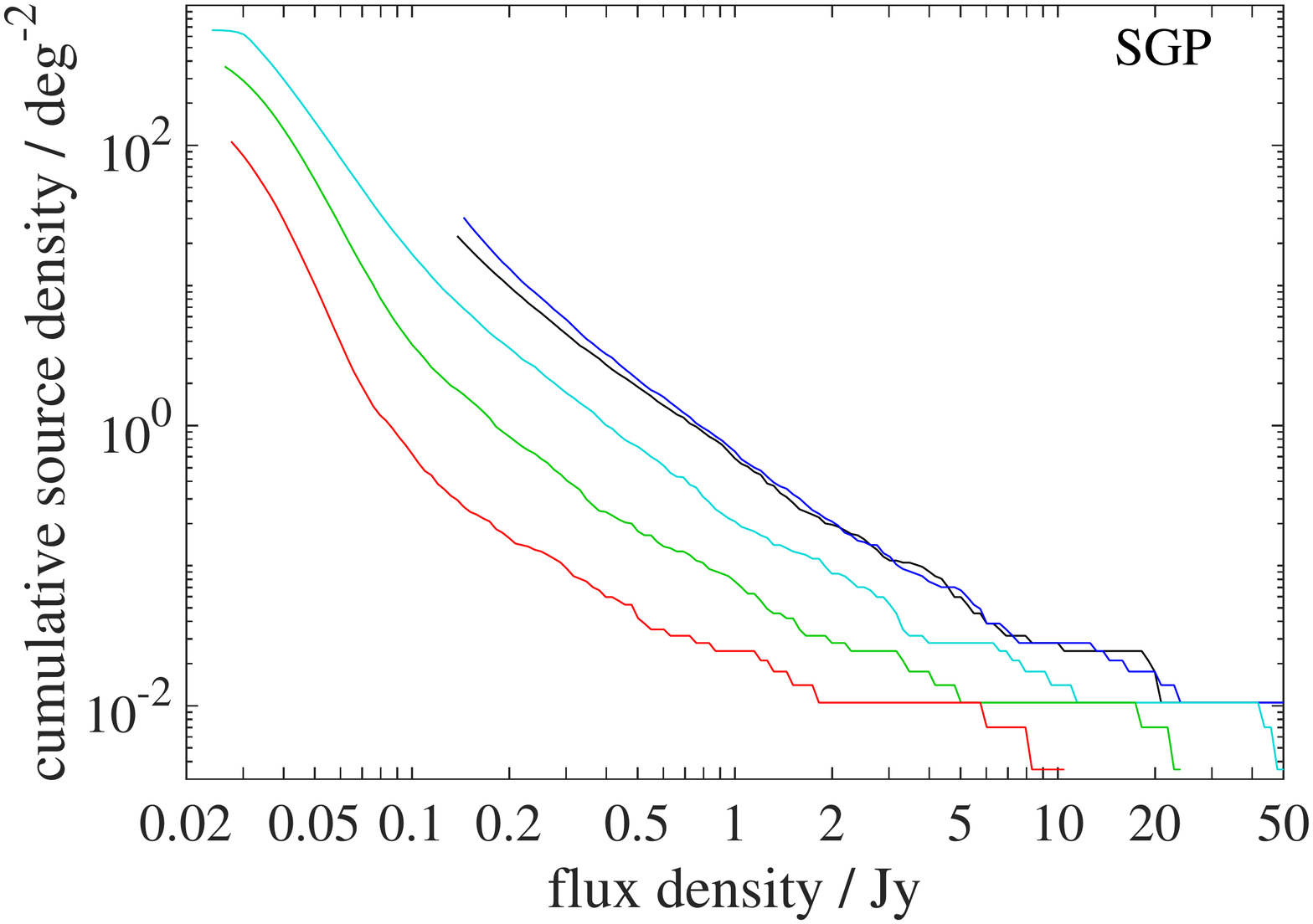}
  \caption{\protect\label{fig_cum_flux} The cumulative number of
    sources as a function of flux density at 100\mic\ (black), 160\mic\ (blue),
    250\mic (cyan), 350\mic\ (green) and 500\mic\ (red). The NGP area is
    shown in the top panel, and the SGP in the bottom panel. The
    counts are plotted only above the limit of 3$\sigma$ in each wave
    band.  }
\end{figure}

The observed number of sources as function of flux density in the PACS and
SPIRE bands is shown in Fig.~\ref{fig_cum_flux}.  Note that this shows
the observed flux in the catalogue, before any corrections are made
for source SED (Section 3.3) or ``flux boosting'' (Section 4.3), which
are necessary before the flux densities are compared with model
predictions. 

\subsection{Positional Accuracy}

V16 carried out extensive simulations to investigate the accuracy of
the H-ATLAS catalogues by injecting artificial sources on to the GAMA
images, and then using {\tt MADX} to detect the sources and measure their
flux densities and positions. The results of these ``in-out''
simulations apply to the NGP and SGP catalogues, which were produced
using almost exactly the same methods.

We investigated the accuracy of the source positions in two ways: (1)
by looking at the positional offsets between the {\it Herschel}
sources and galaxies found on optical images; (2) from the in-out
simulations.  Bourne et al. (2016) and F17 describe the details of the
first method, which takes account of the clustering of the galaxies in
the optical catalogue and the PSF of the {\it Herschel} observations.
Note that astrometric offsets were first calculated using catalogues
from individual {\it Herschel} observations. The astrometry for each
observation was updated before creating the final maps (S17).

In the case of the NGP, we applied this method using the galaxies
found in the SDSS $r$-band images (F17), which thus ultimately ties the
{\it Herschel} positions to the SDSS astrometric frame.   In the case
of the SGP, we used the galaxies found in the VLT Survey Telescope
ATLAS (Shanks et al.  2015), which thus ultimately ties the
astrometry in the SGP to the astrometric frame of this survey.  We
find that the positional error, $\sigma_\mathrm{pos}$, varies from
1.2 to 2.4 arcsec as the signal-to-noise in flux varies from 10 to 5,
with a relationship between positional accuracy and flux density given
by $\sigma_\mathrm{pos} = 2.4 (\mathrm{SNR}/5)^{-0.84}$.  This agrees
well with the errors in the measured positions of the artificial
sources in the in-out simulations (V16). Note that the uncertainty on the
optical positions is typically 0.1 arcseconds, and so is negligible
compared to the {\it Herschel} uncertainties. 
 
\begin{figure*} 
\hspace{10mm}
\includegraphics[scale=0.23]{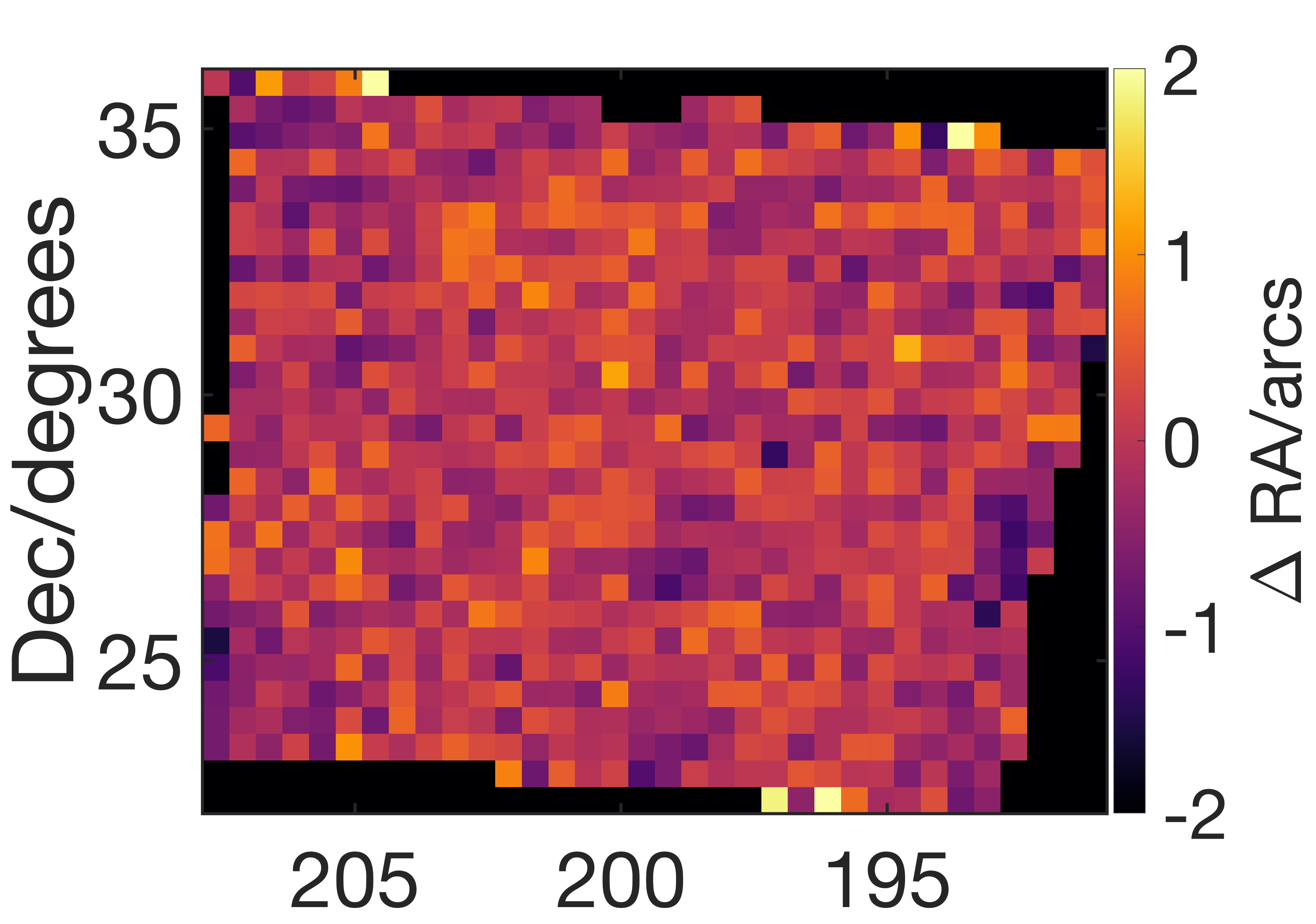}
\hspace{13mm}
\includegraphics[scale=0.23]{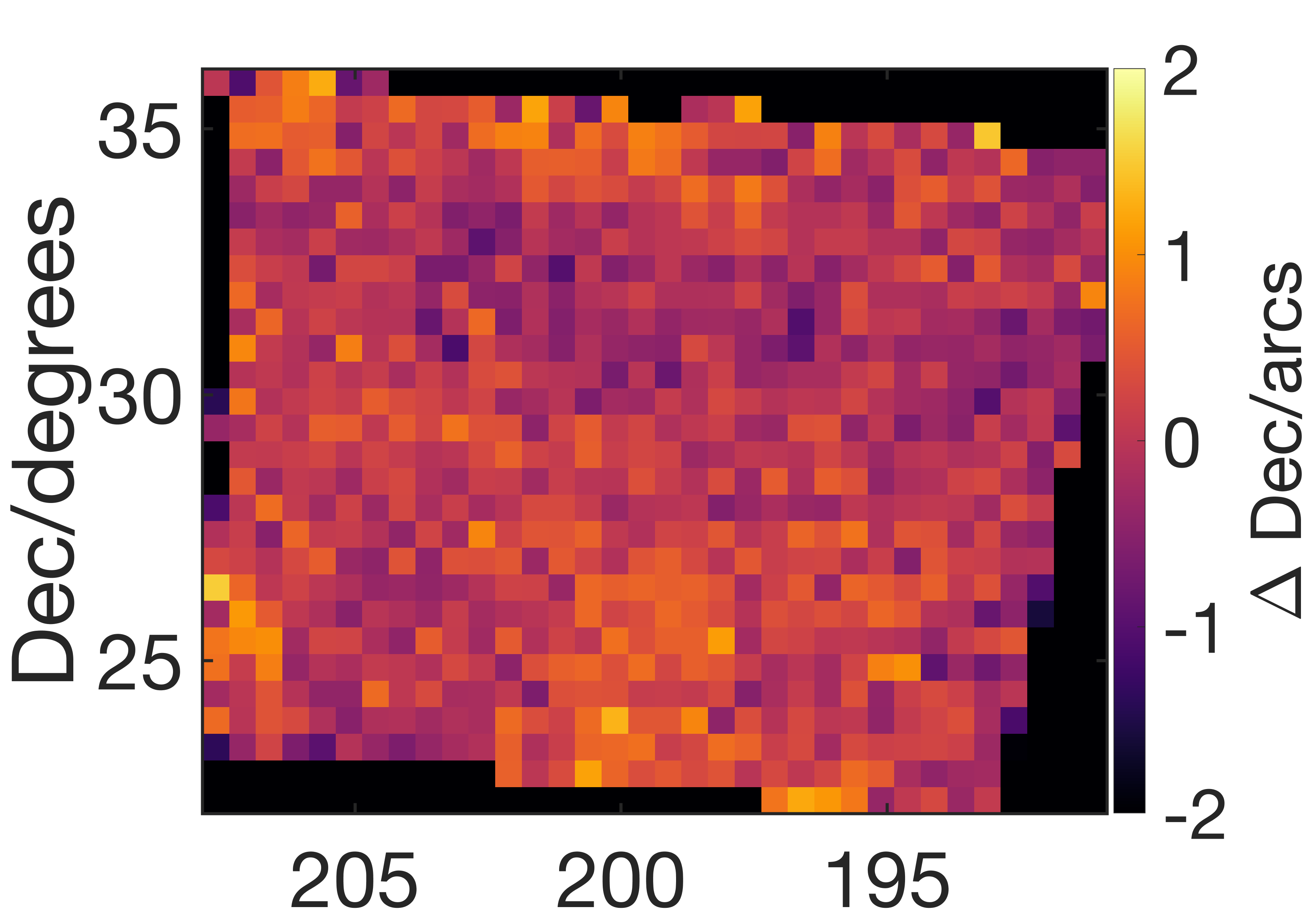}
\hspace{30mm}\\
\includegraphics[scale=0.6,trim={0 86mm 0mm 75mm}, clip]{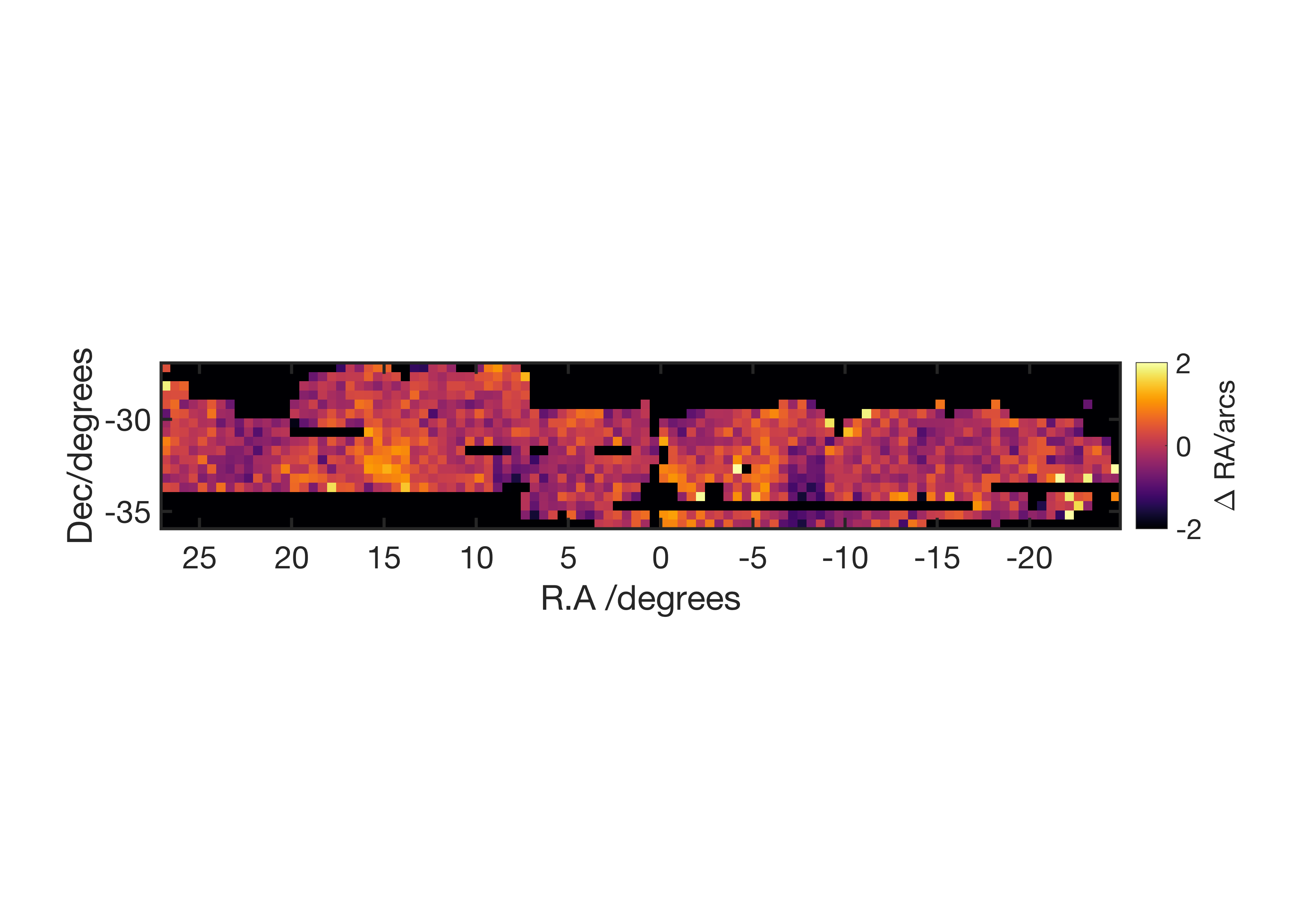}

\includegraphics[scale=0.6,trim={0 60mm 0mm 77mm}, clip]{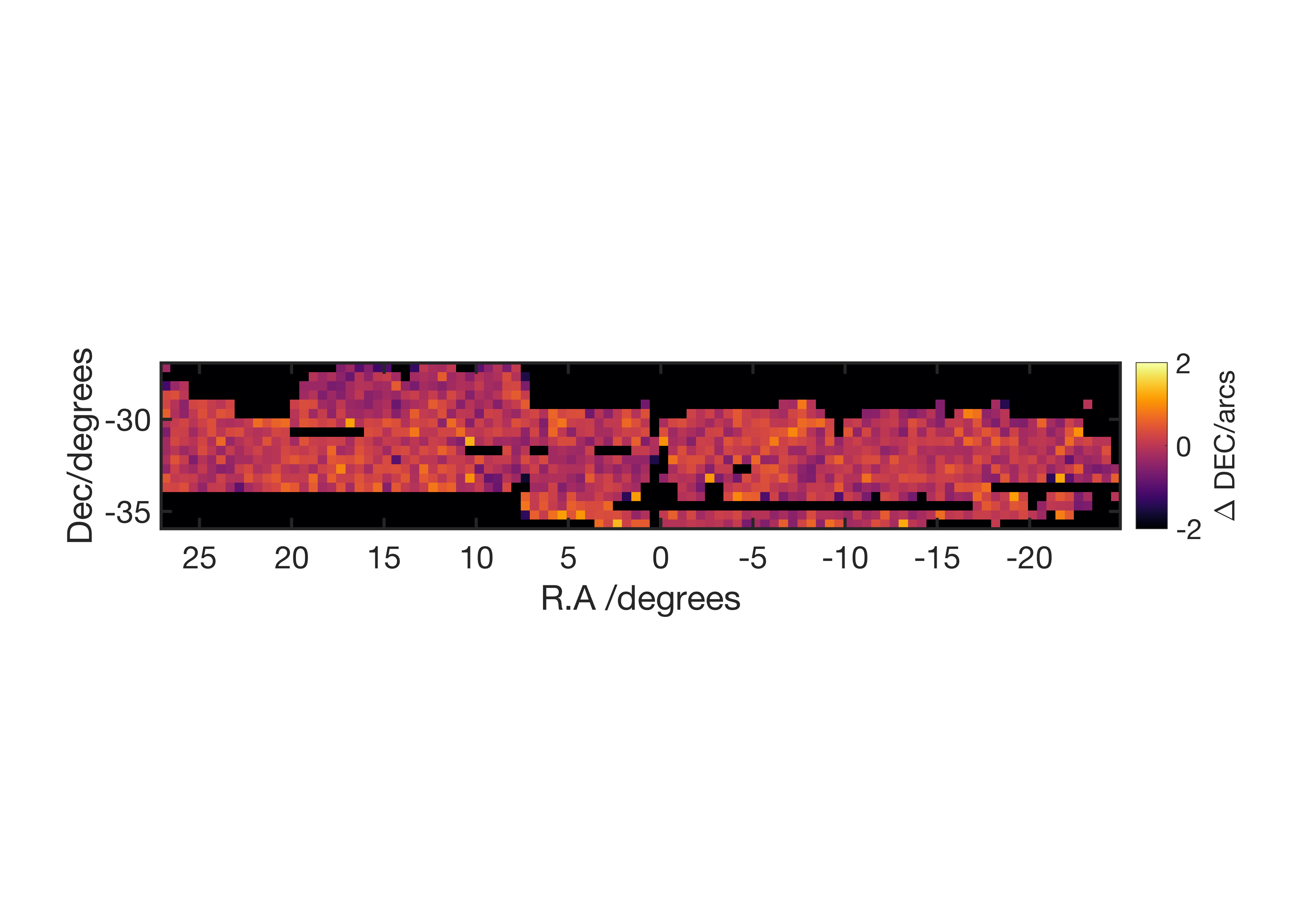}
\caption{\protect\label{fig_pos_errs} The mean positional errors in
  R.A and Dec. averaged in areas 0.5 deg$\times$0.5deg as a
  function of position on the sky for the NGP and SGP
  fields. 
}
\end{figure*}

The mean positional errors as a function of position within the NGP
and SGP fields are shown in Fig.~\ref{fig_pos_errs}. Though there are
hints of systematic variations in different parts of the fields, these
are around 1 arcsec, less than the quoted absolute pointing accuracy of
{\it Herschel} of $\simeq$2 arcsec (Pilbratt et al. 2010).

\subsection{Purity, flux boosting and completeness}

The catalogue is a 4-$\sigma$ catalogue and so we can use Gaussian
statistics to predict the number of sources that will actually be
noise fluctuations; on this basis we expect $\simeq$0.13\% of the
sources in the catalogue to be spurious. However, V16 argue that this
is likely to be a slight overestimate because our errors, while being
good estimates of the errors on the flux measurements, will
underestimate the signal-to-noise of a detection.  To explain in more
detail, our estimate of the confusion noise for a source increases
with increasing source flux, so the noise used to estimate the actual
flux uncertainty is larger than the noise would be for a flux of zero,
as would be appropriate to determine the significance of a
detection. Thus a flux that is four times our quoted error may
correspond to, say, a $4.1$-$\sigma$ detection. In this case our
approximation of contamination from the $4$-$\sigma$ tail of a
Gaussian should be the $4.1$-$\sigma$ tail. In practice the
contamination is so small that the difference is not important and we
have not quantified it.

A major problem in submillimetre surveys, where source confusion is
usually an issue, is flux bias or `flux boosting', in which the
measured flux densities are systematically too high. V16 used the
in-out simulations to quantify this effect in the H-ATLAS. Table 6 in
V16 gives estimates of the flux bias as a function of flux density for
all three SPIRE bands. The table shows that at the 4$\sigma$ detection
flux density, the measured flux densities are on average higher than
the true flux densities by $\simeq$20\%, 5\% and 4\% at 250 $\mu$m,
350 $\mu$m and 500 $\mu$m, respectively. Astronomers interested in
comparing the flux densities in the catalogue with the predictions of
models should be aware of this effect. Following V16, we make no
corrections for this in our catalogue, but Table 6 in V16 can be used
to correct the flux densities for this effect.

Note that the flux limit for a significant PACS detection is much
brighter than the confusion limit, and so PACS fluxes are not affected
by confusion noise. Also the 250-\mic\ noise is so much lower than the
PACS noise, that the 250-\mic\ selection should not introduce any
significant incompleteness in the PACS sample.  The PACS sample should
have completeness and purity as expected for the quoted Gaussian noise
in the flux measurements.

V16 also used the in-out simulations to estimate the completeness of
the survey as a function of measured flux density in all three SPIRE
bands. This is shown in Fig. 21 of V16 and listed in Table 7 of
V16. The completeness at 250\mic\ is 87\% at the 4$\sigma$
detection limit of the survey.

\section{Summary}

We have described the construction of the source catalogues from the
{\it Herschel} survey of fields around the north and south Galactic
poles. This survey which was carried out in five photometric bands --
100, 160, 250, 350 and 500\mic\ -- was part of the {\it Herschel}
Astrophysical Terahertz Large Area Survey (H-ATLAS), a survey of 660
deg$^2$ of the extra-galactic sky. Our source catalogues cover 303
deg$^2$ around the SGP and 177 deg$^2$ around the NGP.

The catalogues contain 118,980 sources for the NGP field and 193,527
sources for the SGP field detected at more than 4$\sigma$ significance
in any of the 250\mic, 350\mic\ or 500\mic\ bands. We present 
photometry in all five bands for each source, including aperture
photometry for sources known to be extended. We discuss all the
practical issues - completeness, reliability, flux boosting, accuracy
of positions, accuracy of flux measurements - necessary to use the
catalogues for astronomical projects.

\section*{Acknowledgements}

PC, LD, HLG, SJM and JSM acknowledge support from the European Research
Council (ERC) in the form of Consolidator Grant {\sc CosmicDust}
(ERC-2014-CoG-647939, PI H\,L\,Gomez).  SJM, LD, NB and RJI acknowledge
support from the ERC in the form of the Advanced Investigator Program,
COSMICISM (ERC-2012-ADG 20120216, PI R.J.Ivison).  EV and SAE
acknowledge funding from the UK Science and Technology Facilities
Council consolidated grant ST/K000926/1.  MS and SAE have received
funding from the European Union Seventh Framework Programme
([FP7/2007-2013] [FP7/2007-2011]) under grant agreement No. 607254.
GDZ acknowledges financial support from ASI/INAF and ASI/University of
Roma Tor Vergata n.$\sim$2014-024-R.1 and n. 2016-24-H.0

The {\it Herschel}-ATLAS is a project carried out using data from {\it
  Herschel}, which is an ESA space observatory with science
instruments provided by European-led Principal Investigator consortia
and with important participation from NASA.

\label{lastpage}

\end{document}